%% file: main.tex
\documentclass[letterpaper, 10 pt, conference]{ieeeconf}  

\IEEEoverridecommandlockouts                              
\usepackage{graphics} 
\usepackage{booktabs}
\usepackage{epsfig} 
\usepackage{mathptmx} 
\usepackage{times} 
\usepackage{amsmath} 
\usepackage{amssymb}  
\usepackage{tikz}
\usepackage{standalone}
\usepackage{bondgraphs}
\usepackage{subcaption}
\usepackage{url}
\usepackage{blox}
\usepackage{breakurl}
\usepackage{blindtext}
\usepackage{rotating}
\usetikzlibrary{shapes,arrows}
\usetikzlibrary{calc, patterns, decorations.pathmorphing, decorations.markings, shapes.geometric, arrows, positioning, angles}

\mathchardef\mhyphen="2D
\newcommand{\freqsize}{0.90}
\newcommand{\freqtext}{0.49}

\newcommand{\symbolsize}{0.92}
\newcommand{\tablesize}{1.00}
\newcommand{\blocksize}{0.90}
\newcommand{\photosize}{0.90}
\newcommand{\modelsize}{0.70}

\usepackage{textcomp}
\usepackage{hyperref}
\usepackage{lipsum}

\title{\LARGE \bf
Modeling and Loop Shaping of Single-Joint Amplification Exoskeleton with Contact Sensing and Series Elastic Actuation
}

\author{Binghan He$^{1}$, Gray C. Thomas, Nicholas Paine and Luis Sentis
\thanks{This work was supported by the U.S. Government and NASA Space Technology Research Fellowship NNX15AQ33H. We thank the members of the Human Centered Robotics Lab, University of Texas at Austin and Apptronik Systems Inc who provided insight and expertise that assisted the research. Authors are with The Departments of Mechanical Engineering (B.H., G.C.T.) and Aerospace Engineering (L.S.), University of Texas at Austin, Austin, TX and with Apptronik Systems Inc (N.P.), Austin, TX. Send correspondence to $^{1}$$\;${\tt\small binghan at utexas dot edu}.
}
}

\newcommand\copyrighttext{%
  \scriptsize 
  Accepted for publication in American Control Conference (ACC)
  \textcopyright 2019 IEEE. Personal use of this material is permitted. Permission from IEEE must be obtained for all other uses, in any current or future media, including reprinting/republishing this material for advertising or promotional purposes, creating new collective works, for resale or redistribution to servers or lists, or reuse of any copyrighted component of this work in other works.
  DOI: \href{https://ieeexplore.ieee.org/abstract/document/8814421}{10.23919/ACC.2019.8814421}
  }
\newcommand\copyrightnotice{%
\begin{tikzpicture}[remember picture,overlay]
\node[anchor=south,yshift=10pt] at (current page.south)
{\fbox{\parbox{\dimexpr\textwidth-\fboxsep-\fboxrule\relax}{\copyrighttext}}};
\end{tikzpicture}%
}

\begin{document}

\maketitle
\thispagestyle{empty}
\pagestyle{empty}
\copyrightnotice

\begin{abstract}
In this paper we consider a class of exoskeletons designed to amplify the strength of humans through feedback of sensed human-robot interactions and actuator forces. We define an amplification error signal based on a reference amplification rate, and design a linear feedback compensator to attenuate this error. Since the human operator is an integral part of the system, we design the compensator to be robust to both a realistic variation in human impedance and a large variation in load impedance. We demonstrate our strategy on a one-degree of freedom amplification exoskeleton connected to a human arm, following a three dimensional matrix of experimentation: slow or fast human motion; light or extreme exoskeleton load; and soft or clenched human arm impedances. We demonstrate that a slightly aggressive controller results in a borderline stable system---but only for soft human musculoeskeletal behavior and a heavy load. This class of exoskeleton systems is interesting because it can both amplify a human's interaction forces ---so long as the human contacts the environment through the exoskeleton--- and attenuate the operator's perception of the exoskeleton's reflected dynamics at frequencies within the bandwidth of the control.
\end{abstract}

\section{Introduction}

\begin{table*}
\caption{List of Symbols}
\centering
\scalebox{\symbolsize}{
\begin{tabular}{r l r l} \toprule

 Symbol & Meaning (Space) & Symbol & Meaning (Space) \\ [.5ex] 
 \midrule
	$K_h, \, B_h, \, M_h$ & Human stiffness, damping and inertia (rotary)  & $\tau_{c}, \, f_c$ & Contact torque (rotary) and force (linear) \\
	$k_h, \, b_h, \, m_h, \, Z_{h}$ & Human stiffness, damping, inertia and impedance (linear) & $\omega_{h}, \, \zeta_{h}$ & Natural frequency and damping ratio of $Z_{h}$\\
	$\bar{M}_e, \, \tilde{M}_e, \, M_e$ & Inertias of exoskeleton, load and combined (rotary) & $\alpha, \, f_{\alpha}$ & Amplification factor and amplification force error (linear) \\
    $\bar{m}_e, \, \tilde{m}_e, \, m_e$ & Inertias of exoskeleton, load and combined (linear) & $\bar{Z}_e, \, \tilde{Z}_e, \, Z_e$ & Impedances of exoskeleton, load and combined (linear)\\
    $Z_{h \mhyphen e}$ & Combined impedance of human and loaded exoskeleton (linear) & $\omega_{h \mhyphen e}, \, \zeta_{h \mhyphen e}$ & Natural frequency and damping ratio of $Z_{h \mhyphen e}$ \\
    $Z_{\alpha h \mhyphen e}$ & Combined impedance of amplified human and loaded exoskeleton (linear) & $\omega_{\alpha h \mhyphen e}, \, \zeta_{\alpha h \mhyphen e}$ & Natural frequency and damping ratio of $Z_{\alpha h \mhyphen e}$\\
    $\theta_{e}, \, x_{e}$ & Exoskeleton joint angle (rotary) and position (linear) & $\tau_{e}, \, f_e$ & Environment torque (rotary) and force (linear) \\
	$k_s, \, Z_s$ & Spring stiffness and impedance (linear) & $\tau_{s}, \, f_s$ & Actuator spring torque (rotary) and force (linear) \\
    $k_{ss}, \, b_{ss}, \, Z_{ss}$ & Virtual spring stiffness, damping and impedance (linear) & $f_{r}$ & Reference spring force (linear) \\
    $b_a, \, m_a, \, Z_a$ & Motor damping, inertia and impedance (linear) & $f_{a}, \, f_{\delta}$ & Motor and disturbance force (linear) \\ 
    $x_a, \, x_d$ & Motor position and position command (linear) & $\tau_{d}, \, f_{d}$ & Desired spring torque (rotary) and force (linear) \\
    $Z_{ss \mhyphen a}$ & Combined impedance of virtual spring and motor (linear) & $\omega_{ss \mhyphen a}, \, \zeta_{ss \mhyphen a}$ & Natural frequency and damping ratio of $Z_{ss \mhyphen a}$ \\
    $P_s , \, P_{\alpha}$ & Transfer functions: from $f_d$ to $f_s$, from $f_d$ to $f_{\alpha}$ & $Q, \, C_s, \, C_{\alpha} $ & DoB filter, spring force and amplification force controller \\
    $u_{h}, \, f_{ext}$ & Biasing force from human and environment (linear) & $C, \, K $ & Extender's position and amplification controller \\ \bottomrule
\end{tabular}
}
\label{symbols}
\end{table*}

Long the purview of science fiction, exoskeletons are quickly becoming a modern reality---augmenting the strength of healthy operators as they walk and interact with the world. A vast literature catalogs the breadth and history of the exoskeleton concept, with survey papers offering disambiguation between such exoskeletons and the orthotic systems designed for medical purposes \cite{DollarHerr2008TRO}, and between  ``parallel-limb exoskeletons for load transfer'' such as our type of system, and several other types that aim to help the human in a different sense \cite{Herr2009JNR} (by reducing the metabolic cost of walking, for example \cite{LeeKimBakerLongKaravasMenardGalianaWalshJNR2018}). Amplification exoskeletons, like the concept of a ``Human Extender'' \cite{Kazerooni1990TSMC}, interact with the world and the human operator at the same time, with the world perceiving a strengthened operator, while the operator in feeling a weakened world and a lighter exoskeleton, all through the feedback action of the device in response to force-sensors embedded at the human--robot interface.

With the human maintaining full control over the motion of the amplification exoskeleton, their primary challenge is not so much autonomy as stable feedback control in the presence of the difficult-to-model human and the uncertain environment. Humans possess naturally adjustable compliance properties which depend on muscle activation \cite{Hogan1984TAC}. The methodology of interaction controller design \cite{ColgateHogan1988IJC} has had success modeling humans as active systems which are passive except for non-state-dependent biases. Robust robot impedance \cite{Hogan1989ICRA} and haptic interface \cite{ColgateBrown1994ICRA,AdamsHannaford1999TRA} controller design strategies have supported and used this model to great effect, while acknowledging its conservatism. The task of amplification exoskeletons, however, is to emulate a reduction in mass. This can be achieved stably if the human model is more precisely known \cite{BuergerHogan2007TRO} than just a passive assumption but is acknowledged to be a challenging problem.

One of the earliest known amplification-oriented exoskeleton
is the hulking machine HARDIMAN I \cite{MakinsonBodineFitck1969Techreport}, which introduced the world to the control challenges of exoskeletons, as it was never safe enough to power on both upper and lower body with an operator. Ref.~\cite{Kazerooni1990TSMC}, much later, conceived of extenders for industrial use with operators controlling much larger machines through force-sensitive interfaces---acknowledging a tradeoff between stability and performance both in linear and robust-nonlinear models \cite{Kazerooni1990TSMC}. Ref.~\cite{KazerooniGuo1993JDSMC} defined a performance criterion for such extenders using a matrix of amplification-levels; a critical frequency, since such amplification cannot be maintained at all frequencies; and introduced a stability filter that allowed the device to ensure robustness to varied operator behavior. However, the later BLEEX exoskeleton from the same lab was not designed in this framework due to practical issues with force sensors \cite{KazerooniRacineHuangSteger2005ICRA} and the discovery of an alternative strategy using high sensitivity cancellation of the natural exoskeleton dynamics---which accomplished an apparent-mass reduction without the force sensors, at the cost of no longer amplifying human-world interaction forces (and extreme sensitivity to the dynamic model) \cite{Kazerooni2005IROS}.

The strategy of measuring a network of force sensors on the human alone, using them to determine human intent, and then using a simulated ideal reaction to this intent as input to a position controller is known as admittance control, and it represents a slight departure from the human extender ideal: accurate reflection of the environmental forces to the human takes a backseat, since there are no force sensors for the environment. This is a very successful paradigm---at least in the absence of environmental contact---it works for giant gantry robots \cite{LecoursStongeGosselin2012ICRA}, complex upper body robots \cite{YuRosen2013TCyb}, the slow-yet-amplifying (full-) body extender \cite{FontanaVertechyMarcheschiSalsedoBergamasco2014RAM}, and the Sarcos-Ratheon exoskeleton described indirectly in their 2014 Patent \cite{JacobsenOlivier2014Patent}---which is implied to be hydraulic, admittance based, and capable of walking. However, when admittance robots interact with semi-rigid environments their interaction forces are determined by the position controller, not the human. A modified admittance scheme with position control implemented via inverse dynamics shares these fundamental limitations \cite{LeeLeeKimJungsooKyoosikHan2014Mechatronics}, and requires either extra force-sensors or modification to the dynamic model to carry load.

A review paper on admittance control \cite{KeeminkKooijStienen2017IJRR} suggests employing feedback based on acceleration when possible. Acceleration feedback (and the use of accelerometers) dominate the control strategy of BLEEX \cite{Kazerooni2005IROS}. And one exoskeleton has used acceleration to reduce the apparent inertia of the operator (rather than merely reducing the apparent mass of the exoskeleton) \cite{KongTomizuka2009TMech}. Acceleration has been proposed as a complete framework for exoskeleton control \cite{BoaventuraBuchli2016BioRob}. But successful as this strategy is, it cannot aid in amplification objectives---since the environment is generally not known in advance. 

In one exoskeleton a simpler strategy was employed which fed  contact forces to motor-current through the motion Jacobian transpose \cite{ZanottoAkiyamaStegallAgrawal2015TRO} (a study not on exoskeleton control but on an exoskeleton's effect on human motion). This strategy is able to amplify the human with respect to both interaction forces and exoskeleton dynamic forces, but the researchers only accomplished a modest amplification (around 2) in their study, and did not discuss the tuning of their controller.

Aiming to accomplish amplification, our exoskeleton hardware is designed to include high performance force controlled series elastic actuators \cite{PaineOhSentis2014TMech} (similar to the force controlled actuation in \cite{KongTomizuka2009TMech}, which was based on \cite{KongBaeTomizuka2009TMech}, which used a disturbance observer on motor position, while ours operates on the spring deflection). Using a series elastic actuator allows us to replicate an amplification error framework reminiscent of \cite{KazerooniGuo1993JDSMC}, but without a force sensor between the exoskeleton and the environment. Instead, we use the spring deflection of the elastic actuators resulting in our scheme to not only attenuate the load but also the exoskeleton dynamics. As in \cite{BuergerHogan2007TRO}, we need to model the human as an uncertain system and check for complementary stability. However, we design our stability filter using the uncertain bode plot between desired force (sent to the SEA) and amplification error. Following the arguments laid out in \cite{BuergerHogan2007TRO}, and the additional examples in \cite{KazerooniSnyder1995IGCD,ColgateBrown1994ICRA}, we employ a single-DOF system to study the simplest possible case of amplification exoskeletons. The non-trivial extension to multi-joint systems remains as future work.

\section{Modeling}

\begin{figure}[!tbp]
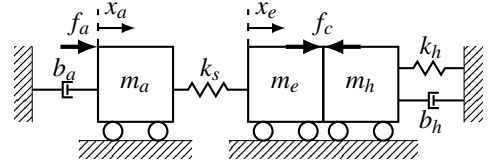

    \centering
    \scalebox{\modelsize}{
    	\includestandalone[width=.5\textwidth]{fig-mass-damper-spring}}
    \caption{A mass-damper-spring model of the human, exoskeleton and SEA.}
    \label{mass-damper-spring}
\end{figure}

\begin{figure*}[!tbp]
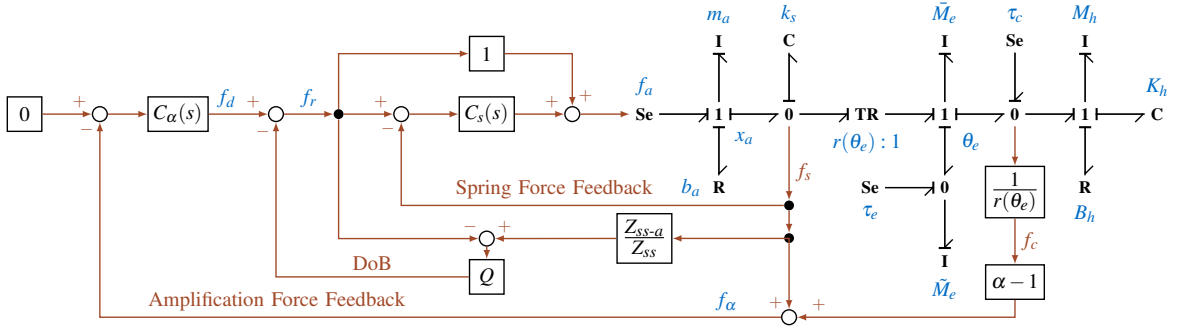

  	\centering
    \scalebox{\blocksize}{
    \includestandalone[width=1.\textwidth]{fig-bond-graph-block-diagram}}
    \caption{Block diagram of spring force feedback, DoB and amplification force feedback. Dynamics of human, exoskeleton and SEA are represented as a bond graph with effort sources $f_a$, $\tau_c$ and $\tau_e$. }
    \label{bond-block}
\end{figure*}

Fig.~\ref{mass-damper-spring} introduces our model of the human--exoskeleton--actuator system, which incorporates a reflection of rotational human elbow motions into a prismatic motion reference frame. 
We define the human forearm moment of inertia $M_{h}$, elbow damping $B_{h}$, elbow stiffness $K_{h}$ and loaded exoskeleton moment of inertia $M_{e} = \bar{M}_{e} + \tilde{M}_e$ (including exoskeleton moment of inertia $\bar{M}_e$ and load moment of inertia $\tilde{M}_e$) as parameters in the rotary joint space. Actuator spring stiffness $k_{s}$, actuator inertia $m_{a}$, and actuator damping $b_{a}$ are parameters in the linear actuator space---with the inertia and damping of the motor rotor being transformed into $m_a$ and $b_a$ in the actuator space by the drive train transmission ratio.

Motor force, $f_{a}$, is the input. Spring force $f_s$, contact torque $\tau_c$---obtained using a six-axis force sensor at the human--robot interface and cast as a torque using the motion Jacobian transpose---, and exoskeleton joint position $\theta_{e}$ are available from sensors. The force from the load to the exoskeleton is not measured but defined as $\tau_e$ in the joint space. Linear versions of rotary position variables are defined based on the joint-angle-dependent inverse kinematics: linear actuator position $x_{a}$, and linear exoskeleton position $x_{e}$. We use the joint-angle-dependent transmission ratio $r (\theta_e) : 1$ to define linear versions of the rotational space parameters: $m_{h} = M_{h} / r(\theta_e) ^ {2}$, $b_{h} = B_{h}/r(\theta_e) ^ {2}$, $k_{h} = K_{h}/r(\theta_e) ^ {2}$, $\bar{m}_{e} = \bar{M}_{e} / r(\theta_e) ^ {2}$, $\tilde{m}_{e} = \tilde{M}_{e} / r(\theta_e) ^ {2}$, $m_e = \bar{m}_{e} + \tilde{m}_{e}$, $f_c = \tau_{c} / r(\theta_e)$ and $f_e = \tau_{e} / r(\theta_e)$. See list of symbols in Tab.~\ref{symbols}.

\setlength{\abovedisplayskip}{4pt}
\setlength{\belowdisplayskip}{4pt}

\subsection{Human-Exoskeleton-Actuator Interaction}

We model the impedance of the human as in \cite{BuergerHogan2007TRO},
\begin{equation} \label{Zh}
Z_{h} = f_{c}/\dot{x}_{e} = m_{h} s + b_{h} + k_{h} s ^ {- 1},
\end{equation}
and the impedance of the unloaded exoskeleton, the load and the loaded exoskeleton,
\begin{equation} \label{Ze}
\begin{aligned}
\bar{Z}_{e} & = (f_{s} - f_{c} - f_{e}) / \dot{x}_{e} = \bar{m}_{e} s, \;
\tilde{Z}_{e} =  f_{e} / \dot{x}_{e} = \tilde{m}_{e} s, \\
Z_{e} & =  (f_{s} - f_{c}) / \dot{x}_{e} = \bar{Z}_{e} + \tilde{Z}_{e} = m_{e} s,
\end{aligned}
\end{equation}
Together in series, 
\begin{equation} \label{xe-fs}
Z_{h \mhyphen e} = Z_{h} + Z_{e} = f_{s} / \dot{x}_{e},
\end{equation}
represents a system in parallel with the elastic actuator.

Considering the impedance of the spring,
\begin{equation}Z_{s} = {f_{s}}/{(\dot{x}_{a} - \dot{x}_{e})} = k_{s} s ^ {-1},\end{equation} and the impedance of the motor, \begin{equation}Z_{a} = {(f_{a} - f_{s})}/{\dot{x}_{a}} = m_{a} s + b_{a},\end{equation} motion of the whole system relates to the required input force,
\begin{equation} \label{xe-fa}
\frac{f_{a}}{\dot{x}_{e}} = Z_{h \mhyphen e} + \frac{(Z_{h \mhyphen e} + Z_{s}) \cdot Z_{a}}{Z_{s}}.
\end{equation}
This provides a human-dependent force-control plant,
\begin{equation} \label{fa-fs}
\frac{f_{s}}{f_{a}} = \frac{Z_{h \mhyphen e} \cdot Z_{s}}{Z_{h \mhyphen e} \cdot Z_{s} + (Z_{h \mhyphen e} + Z_{s}) \cdot Z_{a}}.
\end{equation}

\subsection{Spring Force Control}

Under the spring force control shown in Fig.~\ref{bond-block}'s block diagram bond graph,
\begin{equation} \label{cs}
f_{a} = f_{r} + C_{s} (s) \cdot (f_{r} - f_{s}),
\end{equation}
where $f_{r}$ is the reference spring force and $C_{s} (s)$ is a PD controller. Combining \eqref{fa-fs} and \eqref{cs},
\begin{equation} \label{fss-fs}
\frac{f_{s}}{f_{r}} = \frac{Z_{h \mhyphen e} \cdot Z_{ss}}{Z_{h \mhyphen e} \cdot Z_{ss} + (Z_{h \mhyphen e} + Z_{s}) \cdot Z_{a}},
\end{equation}
where $Z_{ss} = Z_{s} \cdot [1 + C_{s} (s)] = b_{ss} + k_{ss} s ^ {-1}$ is the virtual impedance of spring. By tuning the PD gains of $C_{s} (s)$, the virtual spring stiffness of $k_{ss}$ and the virtual spring damping $b_{ss}$ can be modified.

If $Z_{h \mhyphen e}$ is infinitely large, \eqref{fss-fs} simplifies to,
\begin{equation} \label{fss-fs-norm}
\frac{f_{s}}{f_{r}} = \frac{Z_{ss}}{Z_{ss \mhyphen a}},
\end{equation}
where $Z_{ss \mhyphen a} = Z_{ss} + Z_{a}$ is the combined impedance of the virtual spring and the actuator.

Under the disturbance observer (DoB) of \cite{PaineOhSentis2014TMech}, 
\begin{equation} \label{fss}
f_{r} = f_{d} - [Q \cdot \frac{Z_{ss \mhyphen a}}{Z_{ss}} \cdot f_{s} - Q \cdot f_{r}],
\end{equation}
where $f_{d}$ is the DoB spring force command and $Q$ is a low-pass filter of sufficient order to ensure the observer is causal (2nd order in our analytical model, but 4th order in the low level firmware).

Combining \eqref{fss-fs} and \eqref{fss}, we obtain a transfer function $P_{s}(s)$ from $f_{d}$ to $f_{s}$,
\begin{equation} \label{fd-fs}
P_{s} (s) = \frac{f_{s}}{f_{d}} = \frac{Z_{h \mhyphen e} \cdot Z_{ss}}{Z_{h \mhyphen e} \cdot Z_{ss} + [Z_{h \mhyphen e} + (1 - Q) \cdot Z_{s}] \cdot Z_{a}}.
\end{equation}
By tuning the cut-off frequency of $Q$, \eqref{fss-fs-norm} is (approximately) enforced without an infinitely large $Z_{h \mhyphen e}$---as explained in \cite{PaineOhSentis2014TMech}, this approximation depends on a load inertia lower-bound (in our case, the exoskeleton inertia). 

\section{Loop Shaping}

\begin{figure*}
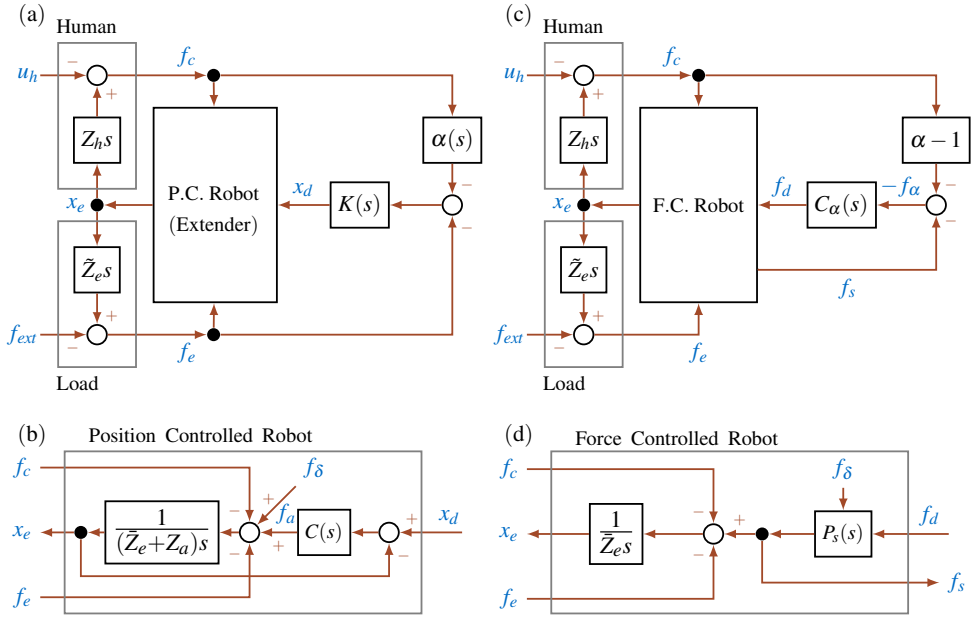

  \begin{minipage}[c]{0.77\textwidth}
    \centering
    \scalebox{1.00}{
    	\includestandalone[width=1.0\textwidth]{fig-extender-and-us}}
  \end{minipage}\hfill
  \begin{minipage}[c]{0.21\textwidth}
    \vspace{4.2cm}
    \caption{Relation between extender concept (a) of \cite{KazerooniGuo1993JDSMC}, which used a position controlled robot (b) as the ``extender'' plant, and our concept of the amplification exoskeleton (c) based on a series elastic actuated, force controlled robot (d).} \label{extender-us}
  \end{minipage}
\end{figure*}

Fig.~\ref{extender-us} shows the amplification exoskeleton concept in comparison with the extender concept of \cite{KazerooniGuo1993JDSMC}. We define the goal of an amplification exoskeleton to be reduction of both environmental and exoskeleton-dynamic load experienced by the user. This differs from an extender, which only seeks to reduce environmental forces---leaving the exoskeleton to be handled by position control. In the extender, drive-train disturbance forces (from stick-slip friction and potentially nonbackdrivable gearing) $f_{\delta}$ are attenuated by position control, while in our system they are attenuated by the DoB. Another important difference is our use of the $f_s$ signal in place of $f_e$; the two are related through
\begin{align}
x_{e} &= \frac{1}{\bar{Z}_{e} s} (f_s - f_e - f_c), \\
f_e + \bar{Z}_{e} s \cdot x_{e} &= f_s - f_c,
\end{align} 
which means that when we replace \cite{KazerooniGuo1993JDSMC}'s feedback of $f_e$ with feedback of $f_s$, we must adjust $\alpha$ by -1 and expect this new controller to attenuate the exoskeleton's own impedance in addition to attenuating environmental forces.

\subsection{Plant of Human Force Amplification}
We define an amplification error signal $f_{\alpha} = (\alpha -1) f_{c} + f_{s}$, and consider the transfer function $P_{\alpha} (s)$ (obtained by combining \eqref{Zh}, \eqref{xe-fs} and \eqref{fd-fs}) from the desired force $f_{d}$ to the error $f_{\alpha}$,
\begin{equation} \label{fd-fe-Q}
\begin{aligned}
P_{\alpha} (s) = \frac{f_{\alpha}}{f_{d}} = \frac{Z_{\alpha h \mhyphen e} \cdot Z_{ss}}{Z_{h \mhyphen e} \cdot Z_{ss} + [Z_{h \mhyphen e} + (1 - Q) \cdot Z_{s}] \cdot Z_{a}},
\end{aligned}
\end{equation}
where $Z_{\alpha h \mhyphen e} = \alpha Z_{h} + Z_{e}$ is the combined impedance of the amplified human and exoskeleton.

Because the cut-off frequency of $Q$ is much larger than the natural frequency of the human system, $\omega_{h} = \sqrt{{k_{h}}/{m_{h}}}$, and the natural frequency of the force control, $\omega_{ss \mhyphen a} = \sqrt{{k_{ss}}/{m_{a}}}$, $P_\alpha(s)$ can be approximated:
\begin{equation} \label{fd-fe}
\begin{aligned}
P_{\alpha} (s) \approx \frac{Z_{\alpha h \mhyphen e} \cdot Z_{ss}}{Z_{h \mhyphen e} \cdot Z_{ss \mhyphen a}}.
\end{aligned}
\end{equation}%

\begin{figure}[!tbp]
    \centering
    \scalebox{\blocksize}{
    	\def\svgwidth{.5\textwidth}
    	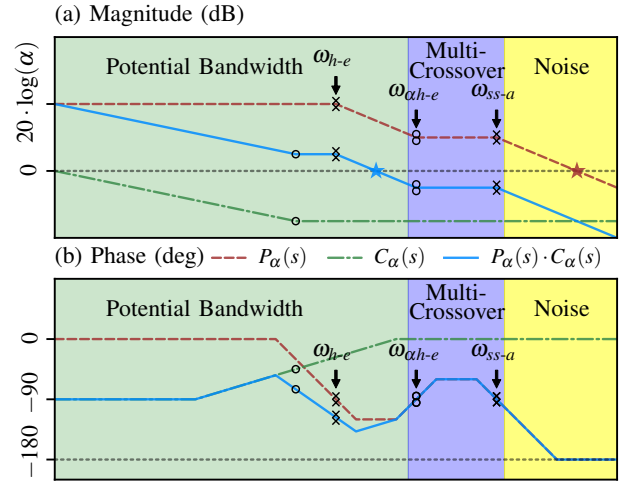
        }
    \caption{Conceptual bode plots with a large $k_{ss} / b_{ss}$ show the plant $P_\alpha(s)$, PI compensator $C(s)$, and compensated open-loop. Regions are color-coded: the green region has practical model-accuracy, the blue region reflects the multi-crossover behavior which makes a compensator design unreliable, and the yellow region is noise dominated in our identification tests, so the model is unreliable. Crosses, circles and stars indicate the poles, zeros and cross-over points.}
    \label{fd-fe-cc}
\end{figure}

As shown in Fig.~\ref{fd-fe-cc}---assuming $k_{ss}/b_{ss}$ is high enough to ignore---$P_{\alpha} (s)$ has a pair of conjugate zeros at $\omega_{\alpha h \mhyphen e} = \sqrt{{k_{h}}/{({m_{e}}/{\alpha} + m_{h})}}$ and two pairs of conjugate poles at $\omega_{h \mhyphen e} = \sqrt{{k_{h}}/{(m_{e} + m_{h})}}$ and at $\omega_{ss \mhyphen a}$. $\omega_{ss \mhyphen a}$ is usually larger than the maximum of $\omega_{h}$. However, even if an actuator has a soft serial spring or a huge motor rotor inertia, increasing the gain of $C_{s} (s)$ will result in an $\omega_{ss \mhyphen a}$ much larger than $\omega_{h \mhyphen e}$ and $\omega_{\alpha h \mhyphen e}$---which avoids the phase drop below $-180 ^{\circ}$ in $P_\alpha(s)$. 

The bode magnitude plot of $P_{\alpha} (s)$ starts from $20 \log (\alpha)$ dB in steady state. If the loop of $P_{\alpha} (s)$ is directly closed, the gain cross-over is decided by the feedback gain---and of course by the shape of $P_{\alpha}(s)$, which varies with the value of $\alpha$, the stiffness and damping of the human, and the environmental impedance. 

When $\alpha$ is set very close to $1$ the actuator does almost no amplification, and $P_\alpha(s)$ approaches the closed loop force tracking behavior of the low-level force controller---in this configuration $\omega_{\alpha h\mhyphen e}$ converges to $\omega_{h\mhyphen e}$ and $P_\alpha(s)$ looks like a second order low pass filter. 

On the other hand, as $\alpha$ increases, $\omega_{\alpha h\mhyphen e}$ travels right, and the gap between $\omega_{h\mhyphen e}$ and $\omega_{\alpha h\mhyphen e}$ widens. This gap behaves like a second order lag compensator---and it corresponds to a phase dip that approaches $-180^\circ$ as the gap widens.

When the human stiffness is very low, both $\omega_{h\mhyphen e}$ and $\omega_{\alpha h\mhyphen e}$ shift to lower frequencies together---unless there is an environmental stiffness. If the human becomes stiffer the two shift higher together as well.

Fig.~\ref{fd-fe-cc} highlights three frequency bands. At the highest frequencies, it should be possible to design controllers, but on our hardware the amount of amplified noise makes them to dangerous to try. In the next highest band it is very easy to gain two extra cross-overs, because the zeros at $\omega_{\alpha h \mhyphen e}$ are under damped. The lowest frequency region is where we will attempt to cross over in this paper.

\subsection{Amplification Force Control}

By adding a proportional gain $k_p$ less than $1$ to $P_{\alpha} (s)$, the cross-over shifts below $\omega_{\alpha h \mhyphen e}$. 
However, It also reduces the low frequency magnitude and increases the closed loop steady state error. 
An integral term can be added to boost the low frequency magnitude while maintaining the same cross-over below $\omega_{\alpha h \mhyphen e}$. We parameterize a simple PI controller transfer function $C_{\alpha} (s)$ (from $- f_{\alpha}$ to $f_d$ in Fig.~\ref{bond-block}) as
\begin{equation} \label{Cc}
C_{\alpha} (s) = k_{p} \cdot \frac{s + z}{s},
\end{equation}
where $z$ is a zero with $k_p \cdot z$ as the integral gain. As a reference, $k_p = \frac{1}{\alpha}$ starts the compensated open-loop plant at a low frequency gain equal to unity.

The amplification tracking is evaluated by comparing $- \frac{f_s}{f_c}$ to $\alpha - 1$. The transfer function from $- (\alpha - 1)f_c$ to $f_s$ is equivalent to closing the loop of $C_{\alpha} (s) P_{s} (s)$.
Because the cut-off frequency of $Q$ is much greater than $\omega_{ss \mhyphen a}$, $C_{\alpha} (s) P_{s} (s)$ can be simplified as 
\begin{equation} \label{PsCc}
C_{\alpha} (s) P_{s} (s) \approx k_{p} \cdot \frac{s + z}{s} \cdot \frac{Z_{ss}}{Z_{ss \mhyphen a}}.
\end{equation}

Notice that $\frac{Z_{ss}}{Z_{ss \mhyphen a}}$ behaves as a low-pass filter with a cut-off frequency at $\omega_{ss \mhyphen a}$. Therefore, $C_{\alpha} (s) P_{s} (s)$ is dominated by $C_{\alpha} (s)$ at low frequency. This allows the amplification tracking to be deterministic despite the uncertainty from $Z_{h \mhyphen e}$.

The dynamic tracking of amplification depends a lot on the location of $z$ in $C_{\alpha} (s)$ and the magnitude of $C_{\alpha} (s) P_{s} (s)$. However, $z$ cannot be arbitrarily large because it allows $C_{\alpha} (s)$ to drop more phase from $P_{\alpha}$ between $\omega_{h \mhyphen e}$ and $\omega_{\alpha h \mhyphen e}$.

\section{System Identification}

The experiments in this section use the elbow-joint exoskeleton testbed (Fig.~\ref{setup}(a)) with a SEA and a contact sensor. The SEA includes a spring of $k_s = 796958 \; N / m$ and a motor of $m_{a} = 250 \; kg$ and $b_a = 4500 \; N s / m$. The aluminum exoskeleton arm has $\bar{M}_{e} = 0.1 \; kg m ^ {2}$ and $r (\theta_e)$ in the range of $[0.005, \, 0.025] \; m$. 

\subsection{Chirp Signal Experiments}\label{chirp}

\begin{figure}[!tbp]
    \centering
    \scalebox{\photosize}{
    	\def\svgwidth{.5\textwidth}
    	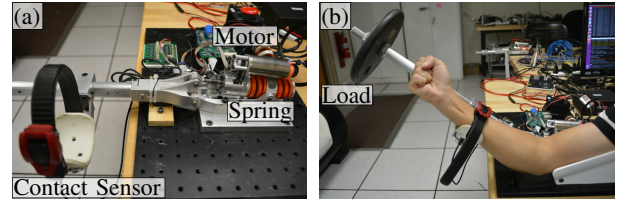
        }
    \caption{An elbow-joint exoskeleton testbed (a) is used for a human model identification experiment (b).}
    \label{setup}
\end{figure}

The human model identification (Fig.~\ref{setup}(b)) includes four chirp signal experiments (Exp.~\ref{chirp}.1-4) with a 27-year old male subject. The subject wears the cuff of the contact sensor and straps his upper arm to a fixed mount. A 300 second, exponential chirp signal is provided as a torque command to the low level force controller. The subject tries to hold the exoskeleton still---with various levels of attempted stiffness---as the exoskeleton vibrates. The exoskeleton does not hit its joint safety hard-stops in these experiments, and operator has access to the system emergency stop at all times.

\begin{figure}[!tbp]
    \centering
    \scalebox{\freqsize}{
    	\def\svgwidth{\freqtext\textwidth}
    	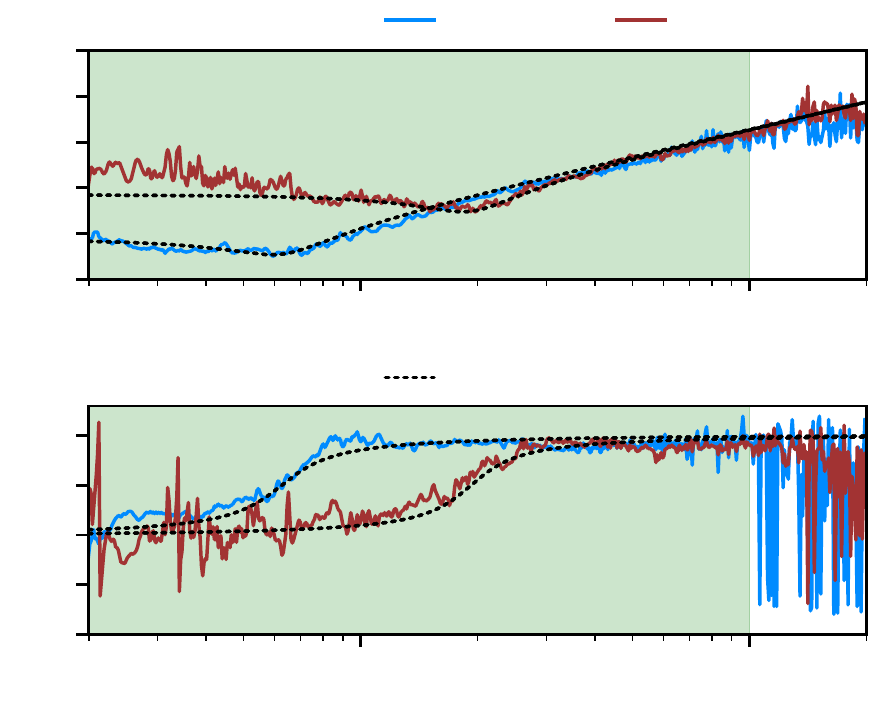
        }
    \caption{Bode plots of $\tau_s / \theta_e\,(s)$, data and fitted models, for Exp.~\ref{chirp}.1 and Exp.~\ref{chirp}.2. Only the green highlighted region is used for identifying $B_{h}$ and $K_{h}$ with regressions.}
    \label{exp-1-2}
\end{figure}
\begin{figure}[!tbp]
    \centering
    \scalebox{\freqsize}{
    	\def\svgwidth{\freqtext\textwidth}
    	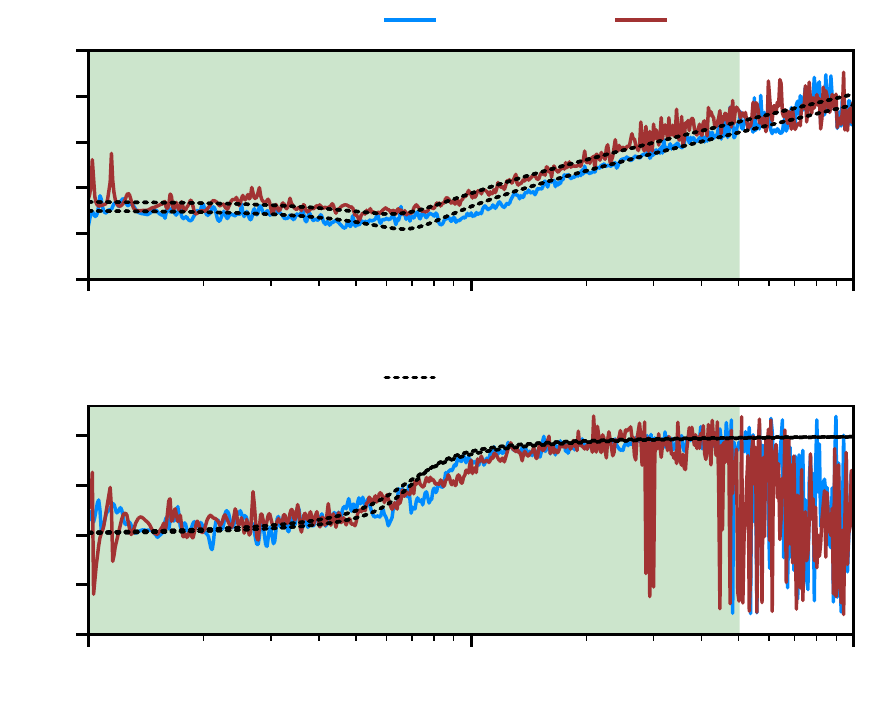
        }
    \caption{Bode plots of $\tau_s/\theta_e\,(s)$, data and fitted models, for Exp.~\ref{chirp}.3 and Exp.~\ref{chirp}.4. Only the green highlighted region is used for identifying $B_{h}$ and $K_{h}$ with regressions.}
    \label{exp-3-4}
\end{figure}
\begin{table}[!]
\caption{Empirical Human Impedance Parameters}
\centering
\scalebox{\tablesize}{
\setlength{\tabcolsep}{4pt}
\begin{tabular}{cccccc} 
\toprule
  Experiment & $K_{h}$$(\frac{N m}{rad})$ & $B_{h}$$(\frac{N m s}{rad})$ & $M_{h}+M_{e}$\scriptsize$(kg m ^ {2})$ & $\zeta_{h \mhyphen e}$ & $|\tau_d|$\scriptsize$(N m)$ \\ [0ex] 
\midrule
 Exp.~\ref{chirp}.$1$ & $7.44$  & $0.56$ & $0.09+0.10$ & 0.24 & 2.0 \\ 
 Exp.~\ref{chirp}.$2$ & $70.11$ & $1.60$ & $0.09+0.10$ & 0.22 & 5.0 \\
 Exp.~\ref{chirp}.$3$ & $31.91$ & $1.84$ & $0.09+0.57$ & 0.20 & 4.0 \\
 Exp.~\ref{chirp}.$4$ & $50.18$ & $4.21$ & $0.09+1.05$ & 0.28 & 4.0 \\ [0ex] 
 \bottomrule
\end{tabular}
}
\label{system-identification}
\end{table}

The time domain data (spring torque $\tau_{s} = f_s \cdot r(\theta_e)$ and angle $\theta_{e}$) from the experiments are used to form a frequency response function representing the human plus the exoskeleton (Figs. \ref{exp-1-2} and \ref{exp-3-4}). The value of $M_{h}$ is identified by the asymptotic behavior at high frequency and the value of $B_h$ and $K_h$ are identified by regressions with identified inertia.

\subsection{Human Joint Stiffness and Damping}

The joint stiffness is decided by contractions in the group of muscles around the joint. By activating a higher level of contractions in flexor and extensor of elbow, the elbow stiffness can be increased \cite{cannon1982mechanical}. Exp.~\ref{chirp}.1 and Exp.~\ref{chirp}.2 identify the range of subject's elbow $K_{h}$ (Fig.~\ref{exp-1-2}). The muscle around the elbow is as relaxed as possible in Exp.~\ref{chirp}.1 and fully tensed in Exp.~\ref{chirp}.2. The subject clenches his hand into a fist to help him achieve a high stiffness and leaves it open for soft behavior. This is also a convenient way to visually distinguish these two types of human behavior in the rest of this paper.

Although the stiffness of a human elbow can possibly go up to $400 \; Nm/rad$, such stiffness is only possible with a perturbation of $40 \; Nm$ \cite{lanman1980movement}. The chirp amplitude $|\tau_d|$ is only $2 \; Nm$ in Exp.~\ref{chirp}.1 and $5 \; Nm$ in Exp.~\ref{chirp}.2 to provide just enough perturbation of torque while the subject can still keep the exoskeleton within the safety joint limits. The results from Exp.~1-2 suggest that $K_h$ varies within the $[7.44, \, 70.11] \; Nm/rad$ range. 

Without additional inertia added to the exoskeleton and load, the human is able to maintain an invariant damping ratio of the arm \cite{perreault2004multijoint}. However, the human is also able to adapt damping and stiffness to compensate the environment dynamics \cite{milner1993compensation}. We added a 5 lb and a 10 lb loads at 18 inches from the joint on the exoskeleton in Exp.~\ref{chirp}.3 and Exp.~\ref{chirp}.4 (Fig.~\ref{exp-3-4}). The results from the four experiments suggest that the human tends to maintain an invariant damping ratio $\zeta_{h \mhyphen e}$ of $Z_{h \mhyphen e}$ when wearing the exoskeleton (Tab.~\ref{system-identification}). Therefore, we model the human as a 1-parameter system. With changing values of $K_h$ and $M_e$, we predict $B_{h}$
\begin{equation} \label{zt}
B_{h} = 2 \zeta_{h \mhyphen e} \sqrt{K_{h} (M_{h} + M_{e})}.
\end{equation}
The identified values of $K_{h}$, $B_{h}$ and $M_{h}$ (Tab.~\ref{system-identification}) suggest the value of $\zeta_{h \mhyphen e}$ of the subject is around $0.23$.

\section{Validation}

Considering the range of $K_{h}$, $m_{e}$ and $r(\theta_e)$, a model of $P_{\alpha} (s)$ with uncertainty can be obtained and used for PI controller design for $\alpha = 10$. In this section, an aggressive controller and a robust controller are implemented to validate the uncertain model and control strategy. $C_s (s)$ is set with $k_{ss} = 2 k_s$ and $b_{ss} = 0.039 k_s$ to achieve a high natural frequency and damping ratio of $Z_{ss \mhyphen a}$. The cut-off frequency of $Q$ is set to be 40 Hz. The video of all the experiments in this section is available at \url{https://www.youtube.com/watch?v=EUHoAEwCfFY}.

\subsection{Aggressive Controller}\label{aggressive}

\begin{figure}[!]
    \centering
    \scalebox{\freqsize}{
    	\def\svgwidth{\freqtext\textwidth}
    	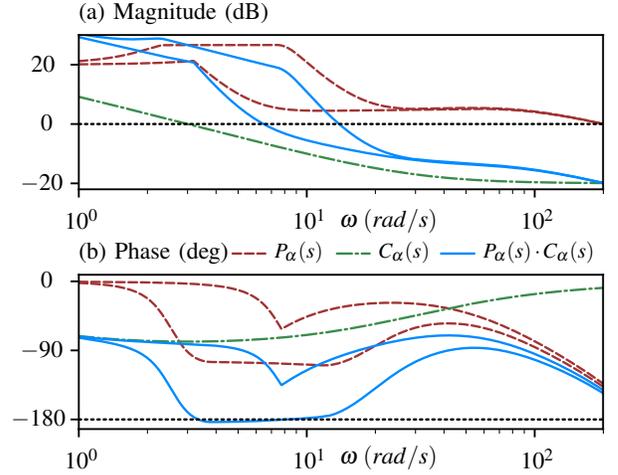
        }
    \caption{Uncertain model of $P_{\alpha} (s)$ is made by 20 interpolations of $K_h$ in the range of $[7.44, \, 70.11] \; N m / rad$ and 5 interpolations of $r(\theta_e)$ in the range of $[0.005, \, 0.025] \; m$. $M_e$ is $1.05 \; kg \cdot m ^ {2}$. Time delay is $6 \; ms$.}
    \label{fd-fe-dd}
\end{figure}

A controller with $k_{p} = 0.1$ and $z = 30$ is implemented with a load of 10 lb on the exoskeleton, the uncertain model suggests it makes the exoskeleton slightly unstable with a low value of $K_{h}$ but fully stable with a high value of $K_{h}$ (Fig.~\ref{fd-fe-dd}).

\begin{figure}[!]
    \centering
    \scalebox{\photosize}{
    	\def\svgwidth{.5\textwidth}
    	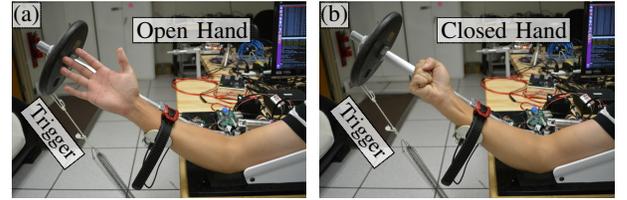
        }
    \caption{Step response with an aggressive controller was triggered by releasing an external spring at the end of the exoskeleton arm. The subject opens (a) (Exp.~\ref{aggressive}.1) and closes (b) (Exp.~\ref{aggressive}.2) his hand to illustrate different levels of muscle co-contraction (and therefore stiffness).}
    \label{setup-agressive}
\end{figure}

\begin{figure}[!]
    \centering
    \scalebox{\freqsize}{
    	\def\svgwidth{\freqtext\textwidth}
    	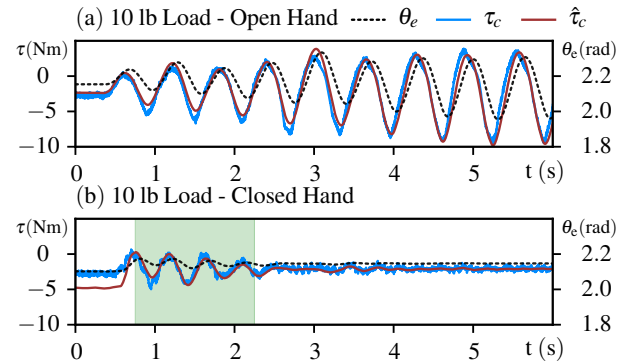
        }
    \caption{Step response results: Exp.~\ref{aggressive}.1 (a) and Exp.~\ref{aggressive}.2 (b). Because motion settles down early in (b), only the green highlighted region is used for estimating $\hat{B}_{h}$ and $\hat{K}_{h}$.}
    \label{fs-fc-p}
\end{figure}
\begin{table}[!]
\caption{Estimated Parameters in Exp.~\ref{aggressive}-series}
\centering
\scalebox{\tablesize}{
\setlength{\tabcolsep}{4pt}
\begin{tabular}{c c c c c c} \toprule
Experiment & Load \scriptsize(lb) & Hand & $\hat{K}_{h}$$(\frac{N m}{rad})$ & $\hat{B}_{h}$$(\frac{N m s}{rad})$ & $\hat{\zeta}_{h \mhyphen e}$ \\ [0ex] 
\midrule 
Exp.~\ref{aggressive}.1 & $10$ & open   & $27.12$ & $2.34$ &  $0.22$ \\
Exp.~\ref{aggressive}.2 & $10$ & closed & $59.34$ & $3.99$ & $0.25$ \\
\bottomrule
\end{tabular}
}
\label{fs-fc-p-table}
\end{table}

To validate this prediction, the subject wearing the exoskeleton with the controller either opens his hand---for low values of $K_{h}$---or tightly closes his hand---for high stiffness $K_{h}$---in two separate experiments: Exp.~\ref{aggressive}.1 and Exp.~\ref{aggressive}.2 respectively. A spring with one end connected on the ground and the other end hanging on the end of the exoskeleton is released at the beginning of each experiment---a step input in external force (Fig.~\ref{setup-agressive}).

The experiment results (Fig.~\ref{fs-fc-p}) show that the exoskeleton joint oscillates with a slowly increasing amplitude with open hand and oscillates but settles down in 2 seconds with closed hand. By using the data of $\tau_c$, $\theta_e$ and $\dot{\theta}_e$, a linear regression is built to identify a simplified human model with only damping $\hat{B}_{h}$ and stiffness $\hat{K}_{h}$ (Tab.~\ref{fs-fc-p-table}). The value of $\hat{K}_h$ verifies that human maintains a much higher stiffness with closed hand than open hand. The estimated contact torques $\hat{\tau}_c = \hat{B}_{h} \dot{\theta}_e + \hat{K}_{h} \theta_e$ are well matched to the measured $\tau_c$ (Fig.~\ref{fs-fc-p})---which confirms that the system is oscillating despite a passive, spring-damper-like, human behavior. The estimated damping ratio $\hat{\zeta}_{h \mhyphen e} = \hat{B}_{h}/(2 \sqrt{\hat{K}_{h}M_{e}})$ for open hand is slightly lower than 0.23 and makes the system more unstable than the prediction. 

\subsection{Robust Controller}\label{robust}

\begin{figure}[!]
    \centering
    \scalebox{\freqsize}{
    	\def\svgwidth{\freqtext\textwidth}
    	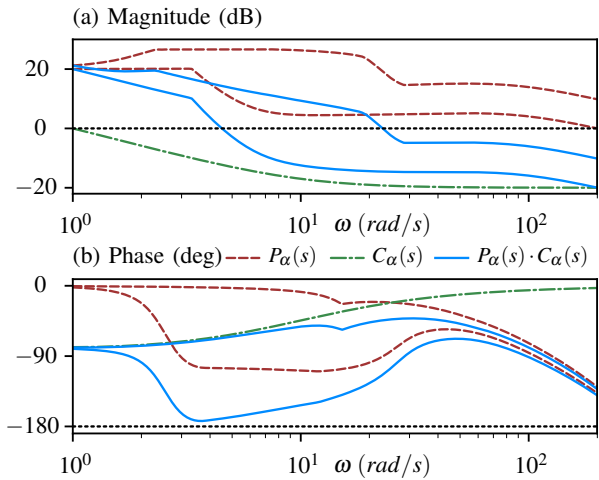
        }
    \caption{Uncertain model of $P_{\alpha} (s)$ is made by 20 interpolations of $K_h$ in the range of $[7.44, \,70.11] \; N m / rad$, 20 interpolations of $M_e$ in the range of $[0.1, \, 1.05] \; kg \cdot m ^ {2}$, and 5 interpolations of $r(\theta_e)$ in the range of $[0.005, \, 0.025] \; m$. Time delay is $6 \; ms$.}
    \label{fd-fe-aa}
\end{figure}

We also implemented a controller with $k_p = 0.1$ and $z = 10$ for improved robustness to parameter variation in the exoskeleton load, and human stiffness (and damping). The uncertain model suggests it maintains a phase margin no less than $10 ^ {\circ}$ for permissible values of $K_{h}$, $M_e$, and $r(\theta_e)$ (Fig.~\ref{fd-fe-aa}).

To validate the expectation, the experiments include two different $M_{e}$ (with 0 and 10 lb load) and two different $K_{h}$ (open and closed hands). The subject generates motion with 0.1 Hz trapezoid-like wave for steady state tests and 1 Hz sinusoid-like wave for dynamic tests. We refer to these eight experiments as the Exp.~\ref{robust}-series.

The integrator in the controller is implemented as a pole $p = 0.01$ to numerically integrate $f_{\alpha}$. Because the amplification tracking relies mostly on $C_{\alpha} (s)$, a static gain of 8.18 and a 1 Hz dynamic gain of $1.42$ with phase shift of $-56.94 ^ {\circ}$ are expected for $- f_s / f_c$ (or $- \tau_s / \tau_c$ in joint space ).

\begin{figure*}
  \begin{minipage}[c]{0.81\textwidth}
    \centering
    \scalebox{.79}{
    	\def\svgwidth{1.15\textwidth}
        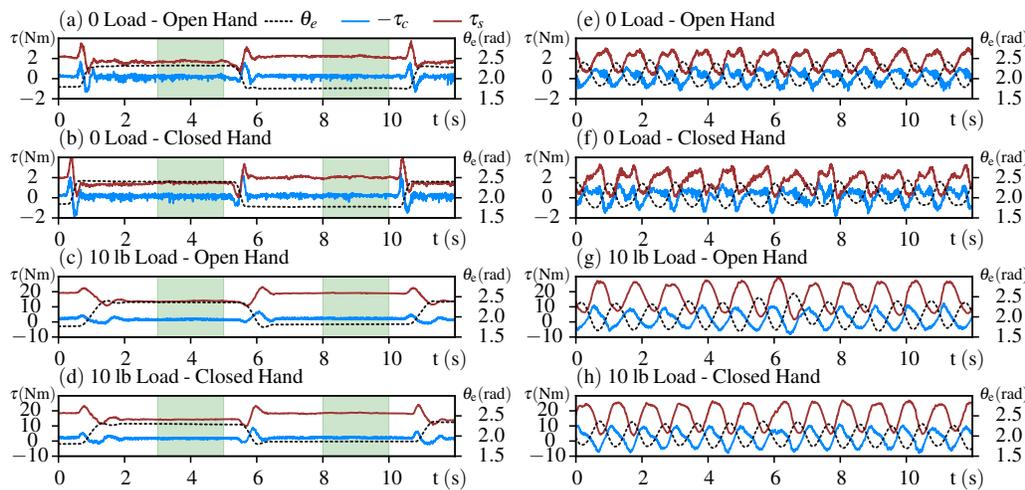
        }
  \end{minipage}\hfill
  \begin{minipage}[c]{0.19\textwidth}
    \vspace{2.2cm}
    \caption{Results from the eight Exp.~\ref{robust}-series tests. Steady state tests (a)-(d) and dynamic tests (e)-(h) are processed from the subject. Only data in the green highlighted regions are used for calculating the static gain.} \label{fs-fc-s-f}
  \end{minipage}
\end{figure*}

The experiment results (Fig.~\ref{fs-fc-s-f}) verify that the exoskeleton is stable with all eight settings. The values of gain and phase shift are also close to the expected values (Tab.~\ref{fs-fc-s-f-table}). The results show little influence on amplification tracking from the variation of $Z_{h}$ and $M_{e}$, as expected.

\section{Discussion}

Amplification exoskeletons, seeking to amplify the strength of the human as it interacts both with the environment and the inertia of the robot itself, represent a different strategy of human augmentation compared to the extender concept of \cite{KazerooniGuo1993JDSMC}. Series elastic actuators fit into the structure of amplification exoskeleton by incorporating a direct measurement of actuator force through spring deflection and an accurate actuator force control (with a disturbance observer).

We showcase a proportional-integral amplification force controller which aims to stably achieve high steady-state amplification tracking performance and a large amplification factor. Since an aggressive controller (with a high zero frequency $z$) causes stability problem in the case of low human joint stiffness, our results with a robust controller corroborate \cite{BuergerHogan2007TRO} in demonstrating that the variability in human stiffness must be accounted for in designing the controller. Our control strategy leaves exoskeleton designers with a trade-off between high amplification factors, high bandwidth of the amplification error attenuation function, and accurate knowledge of the human stiffness limits.

\begin{table}
\caption{Observed Amplification in Exp.~\ref{robust}-series}
\centering
\scalebox{\tablesize}{
\setlength{\tabcolsep}{4pt}
\begin{tabular}{c c c c c} \toprule
 Load \scriptsize(lb) & Hand & \small$|- \frac{\tau_s}{\tau_c}|$ \scriptsize(static) & \small$|- \frac{\tau_s}{\tau_c}|$ \scriptsize(1Hz) & \small$\angle- \frac{\tau_s}{\tau_c}$ \scriptsize(1Hz) \\ [.5ex] 
 \midrule
 0 & open   & $8.15$ & $1.48$ & $-53.55 ^ {\circ}$ \\ 
 0 & closed & $8.05$ & $1.49$ & $-53.68 ^ {\circ}$ \\
10 & open   & $8.13$ & $1.58$ & $-53.39 ^ {\circ}$ \\
10 & closed & $8.18$ & $1.46$ & $-50.75 ^ {\circ}$ \\
\bottomrule
\end{tabular}
}
\label{fs-fc-s-f-table}
\end{table}

Our strategy can potentially be extended to multi-DOF cases by projecting the human-exoskeleton interaction forces measured from the contact sensors to each joint space and developing joint level amplification force controllers accordingly. Actuator force saturation may appear on amplification exoskeletons with other human joints (especially those at the lower limbs) where higher maximum joint torques can exist. This saturation problem can be considered as a reduction of the actual amplification factor and certainly reduces the bandwidth and crossover frequency of force amplification. Nevertheless, a decrease in crossover frequency (as discussed on Fig.~\ref{fd-fe-cc}) does not create an additional stability issue for our proportional-integral controller.

\addtolength{\textheight}{-0cm}   








\bibliographystyle{IEEEtran}
\bibliography{main}


\end{document}

%% file: fig-mass-damper-spring.tex
\begin{tikzpicture}
\tikzstyle{spring}=[thick,decorate,decoration={zigzag,pre length=0.2cm,post length=0.2cm,segment length=6}]
\tikzstyle{damper}=[thick,decoration={markings,  
  mark connection node=dmp,
  mark=at position 0.5 with 
  {
    \node (dmp) [thick,inner sep=0pt,transform shape,rotate=-90,minimum width= 5pt,minimum height=1pt,draw=none] {};
    \draw [thick] ($(dmp.north east)+(2pt,0)$) -- (dmp.south east) -- (dmp.south west) -- ($(dmp.north west)+(2pt,0)$);
    \draw [thick] ($(dmp.north)+(0,-2pt)$) -- ($(dmp.north)+(0,2pt)$);
  }
}, decorate]
\tikzstyle{ground}=[fill,pattern=north east lines,draw=none,minimum width=0.75cm,minimum height=0.15cm]

\begin{scope}[xshift=5cm]
\node [draw,outer sep=0pt,thick] (M1) [minimum width=1cm, minimum height=1.0cm] { $m_a$};

\node [draw,outer sep=0pt,thick] (M2) [minimum width=1cm, minimum height=1.0cm, xshift = 2.cm] { $m_e$};

\node [draw,outer sep=0pt,thick] (M3) [minimum width=1cm, minimum height=1.0cm, xshift = 3.cm] { $m_h$};

\node (ground1) [ground,anchor=north, yshift=-0.25cm, minimum width=1.5cm] at (M1.south) {};

\draw (ground1.north east) -- (ground1.north west);

\node (ground2) [ground,anchor=north, yshift=-0.25cm, minimum width=2.5cm] at (M2.south east) {};

\draw (ground2.north east) -- (ground2.north west);

\draw [thick] (M1.south west) ++ (0.2cm,-0.125cm) circle (0.125cm)  (M1.south east) ++ (-0.2cm,-0.125cm) circle (0.125cm);

\draw [thick] (M2.south west) ++ (0.2cm,-0.125cm) circle (0.125cm)  (M2.south east) ++ (-0.2cm,-0.125cm) circle (0.125cm);

\draw [thick] (M3.south west) ++ (0.2cm,-0.125cm) circle (0.125cm)  (M3.south east) ++ (-0.2cm,-0.125cm) circle (0.125cm);

\node (wall1) [ground, rotate=-90, minimum width = 1cm, yshift = - 1.5cm] {};

\node (wall2) [ground, rotate=-90, minimum width = 1cm, yshift =   4.5cm] {};

\draw (wall1.north east) -- (wall1.north west);

\draw (wall2.south east) -- (wall2.south west);


\draw [spring] ($(M1.north east)!(wall1.60)!(M1.south east)$) -- ($(M2.north west)!(wall1.60)!(M2.south west)$) node [midway,above] { $k_s$};

\draw [damper] ($(wall1.north east)!(wall1.60)!(wall1.north west)$) -- ($(M1.north west)!(wall1.60)!(M1.south west)$) node [midway,above] { $b_a$};

\draw [spring] ($(M3.north east)!(wall2.150)!(M3.south east)$) -- ($(wall2.south east)!(wall2.150)!(wall2.south west)$) node [midway,above] { $k_h$};

\draw [damper] ($(M3.north east)!(wall2.30)!(M3.south east)$) -- ($(wall2.south east)!(wall2.30)!(wall2.south west)$) node [midway,below] { $b_h$};


\draw [-latex,ultra thick] ($(M2.north east)$) ++ (-.5cm,-0.cm) -- +( .5cm,0cm) node [right,above] { $f_c$};

\draw [-latex,ultra thick] ($(M3.north west)$) ++ ( .5cm,-0.cm) -- +(-.5cm,0cm) node [midway,above] {};

\draw [-latex,ultra thick] ($(M1.north west)$) ++ (-.5cm,-0.cm) -- +( .5cm,0cm) node [midway,above] { $f_a$};


 \draw[thick, dashed] ($(M1.north west)$) -- ($(M1.north west) + (0,.5)$);
 \draw[thick, dashed] ($(M2.north west)$) -- ($(M2.north west) + (0,.5)$);
 \draw[thick, -latex] ($(M2.north west) + (0,0.25)$) -- ($(M2.north west) + (0.5,0.25)$) node [midway, above] { $x_e$};
 \draw[thick, -latex] ($(M1.north west) + (0,0.25)$) -- ($(M1.north west) + (0.5,0.25)$) node [midway, above] { $x_a$};

\end{scope}
\end{tikzpicture}

%% file: fig-bond-graph-block-diagram.tex
\definecolor{blue}{RGB}{0, 108, 198}
\tikzstyle{block}=[minimum width=7mm,minimum height=7mm,draw,thick]
\tikzstyle{plusminus}=[circle,thick,draw,inner sep=0pt,minimum size=3mm]
\tikzstyle{splitter}=[circle,fill,minimum size=2mm,inner sep=0pt]
\tikzstyle{signal} = [-latex, thick, color={rgb,255:red,162; green,51; blue,51}]
\begin{bondgraph}
\begin{scope}[every node/.style={bgelement}]
\node[label=above: \large $f_a$] at (0,0) (S1) {Se};
\node[label=south east: \large $x_a$,right=1. of S1] (xa) {1};
\node[label=left : \large $b_a$,below=1. of xa] (ba) {R};
\node[label=above: \large $m_a$,above=1. of xa] (ma) {I};
\node[right=1. of xa] (fs) {0};
\node[label=above: \large $k_s$,above=1. of fs] (ks) {C};
\node[label=below: \large $r(\theta_e):1$,right=1. of fs] (tr) {TR};
\node[label=south east: \large $\theta_e$,right=1. of tr] (xe) {1};
\node[below=1. of xe] (fe) {0};
\node[label=below: \large $\tilde{M}_e$,below=1. of fe] (ml) {I};
\node[label=below: \large $\tau_e$,left =1. of fe] (S3) {Se};
\node[label=above: \large $\bar{M}_e$,above=1. of xe] (me) {I};
\node[right=1. of xe] (fc) {0};
\node[label=above: \large $\tau_c$,above=1. of fc] (S2) {Se};
\node[right=1. of fc] (xh) {1};
\node[label=below: \large $B_h$,below=1. of xh] (bh) {R};
\node[label=above: \large $M_h$,above=1. of xh] (mh) {I};
\node[label=above: \large $K_h$,right=1. of xh] (kh) {C};
\end{scope}
\draw
(S1) edge[e_out] (xa)
(xa) edge[f_out] (ba)
(xa) edge[e_out] (ma)
(xa) edge[f_out] (fs)
(fs) edge[f_out] (ks)
(fs) edge[e_out] (tr)
(tr) edge[e_out] (xe)
(xe) edge[e_out] (me)
(xe) edge[f_out] (fc)
(xe) edge[f_out] (fe)
(fe) edge[e_out] (ml)
(S2) edge[e_out] (fc)
(S3) edge[e_out] (fe)
(fc) edge[e_out] (xh)
(xh) edge[f_out] (bh)
(xh) edge[f_out] (kh)
(xh) edge[e_out] (mh);
\node[splitter,below=1.5 of fs] (s2) {};
\node[splitter,below=.45 of s2] (s3) {};
\node[plusminus,left=1. of S1] (p1) {};
\node[block,left=1. of p1] (Cs) {\large $C_{s} (s)$};
\node[block,above=.5 of Cs] (ff) {\large $1$};
\node[plusminus,left=1. of Cs] (p2) {};
\node[splitter,left=1. of p2] (s1) {};
\node[plusminus,left=1. of s1] (p3) {};
\node[block,below=1.76 of S1] (norm) {\LARGE $\frac{Z_{ss \mhyphen a}}{Z_{ss}}$};
\node[plusminus,below=2. of Cs] (p4) {};
\node[block,below=.25 of p4] (Q) {\large $Q$};
\node[block,left =1.20 of p3] (Cc) {\large $C_{\alpha} (s)$};
\node[plusminus,left=.8 of Cc] (p5) {};
\node[block,below=.8    of fc] (invr) {\LARGE $\frac{1}{r(\theta_e)}$};
\node[block,below=2.8   of fc] (alpha) {\large $\alpha - 1$};
\node[block,left=1. of p5] (setpoint) {\large $0$};
\node[plusminus,below=2. of s2] (p6) {};
\begin{scope}[every path/.style={signal}]
\draw (setpoint) -- node[near end,above]{$+$} (p5);
\draw (p2) -- (Cs);
\draw (s1) |- (ff);
\draw (ff) -| node[pos=.9,right]{$+$} (p1);
\draw (s1) -- node[near end,above]{$+$} (p2);
\draw (Cs) -- node[near end,above]{$+$} (p1);
\draw (p1) -- (S1);
\draw (fs) -- node[pos=.625,right]{\large $f_s$} (s2);
\draw (s2) -| node[pos=.3 ,above]{\large \rm{Spring Force Feedback}} node[pos=.96,left]{$-$} (p2);
\draw (s2) -- (s3);
\draw (s3) |- (norm);
\draw (p3) -- node[pos=0.5 ,above]{\textcolor{blue}{\large $f_{r}$}}(s1);
\draw (norm) -- node[pos=0.92,above]{$+$}(p4);
\draw (s1)   |- node[pos=0.97,above]{$-$}(p4);
\draw (p4)   -- (Q);
\draw (Q)    -| node[pos=0.25,above]{\large \rm{DoB}} node[pos=0.97,left ]{$-$}(p3);
\draw (Cc)   -- node[pos= .3 ,above]{\textcolor{blue}{\large $f_d$}} node[near end,above]{$+$}(p3);
\draw (p5)   -- (Cc);
\draw (fc) -- (invr);
\draw (invr) -- node[pos= .5 ,right]{\large $f_c$} (alpha);
\draw (alpha) |- node[pos= .95,above]{$+$} (p6);
\draw (s3) -- node[pos= .9 ,left ]{$+$} (p6);
\draw (p6) -| node[pos= .37,above]{\large \rm{Amplification Force Feedback}} node[pos= .04,above]{\textcolor{blue}{\large $f_{\alpha}$}} node[pos= .98,left ]{$-$} (p5);
\end{scope}
\end{bondgraph}

%% file: fig-extender-and-us.tex
\definecolor{blue}{RGB}{0, 108, 198}
\tikzstyle{block}=[minimum width=7mm,minimum height=7mm,draw,thick]
\tikzstyle{plusminus}=[circle,thick,draw,inner sep=0pt,minimum size=3mm]
\tikzstyle{splitter}=[circle,fill,minimum size=2mm,inner sep=0pt]
\tikzstyle{dot}=[circle,minimum size=0mm,inner sep=0pt]
\tikzstyle{signal} = [-latex, thick, color={rgb,255:red,162; green,51; blue,51}]
\begin{bondgraph}

\begin{scope}[xshift=0cm, yshift=-.2cm]
\node[plusminus] (fh) {};
\node[splitter, right=1.525 of fh] (s1) {};
\node[block, below of=fh] (Hs) {$Z_{h} s$};
\node[splitter, below of=Hs] (sp1) {};
\node[block, right=.75 of sp1, minimum height=3cm, minimum width=1.9cm] (Gs) {};
\node[dot, below=1.0 of s1] (Ex) {\small $\rm{P.C. \; Robot}$};
\node[dot, below=1.5 of s1] (Ex) {\small $\rm{(Extender)}$};

\node[block, right= .8 of Gs] (Ks) { $K(s)$};
\node[plusminus, right= .8 of Ks] (err) {};
\node[block, above of=err] (alphas) { $\alpha(s)$};

\node[block, below of=sp1] (Es) {$\tilde{Z}_{e} s$};

\node[plusminus, below of=Es] (fe) {};
\node[splitter, right=1.525 of fe] (s2) {};
\node[left of=fh] (uh) {};
\node[left of=fe] (ue) {};

\node[left= .25 of uh, above=.5 of uh] (parta) {$\rm{(a)}$};
\begin{scope}[every path/.style={signal}]
\draw (Es) -- node[near end, right]{\scriptsize $+$} (fe);
\draw (fe) -- node[pos = 0.8, below]{\textcolor{blue}{$f_e$}} (s2);
\draw (s2) -- (Gs);
\draw (s2) -| node[pos = 0.95, right]{\scriptsize $-$} (err);
\draw (Gs) -- node[pos= 1.2, left]{\textcolor{blue}{$x_e$}} (sp1);
\draw (sp1) -- (Hs);
\draw (sp1) -- (Es);
\draw (Hs) -- node[near end, right]{\scriptsize $+$} (fh);
\draw (fh) -- node[pos = 0.8, above]{\textcolor{blue}{$f_c$}} (s1);
\draw (s1) -- (Gs);
\draw (s1) -| (alphas);
\draw (alphas) -- node[pos = 0.7, right]{\scriptsize $-$} (err);
\draw (err) -- (Ks);
\draw (Ks) -- node[pos=0.5, above]{\textcolor{blue}{$x_{d}$}} (Gs);
\draw (uh) -- node[left=.25 of uh]{\textcolor{blue}{$u_h$}}     node[pos = 0.7, above]{\scriptsize $-$} (fh);
\draw (ue) -- node[left=.25 of ue]{\textcolor{blue}{$f_{ext}$}} node[pos = 0.7, below]{\scriptsize $-$} (fe);
\end{scope}
\draw [color=gray,thick](-.6,.5) rectangle (.6,-1.75);
\node at (-.75,.75) [above=5mm, right=0mm] {\small $\rm{Human}$};
\draw [color=gray,thick](-.6,-2.25) rectangle (.6,-4.5);
\node at (-.75,-4.75) [above=5mm, right=0mm] {\small $\rm{Load}$};
\end{scope}

\begin{scope}[xshift=-1cm, yshift=-5.75cm, on grid]
\node[] (partb) {$\rm{(b)}$};
\node[below=.5 of partb] (fh2) {};
\node[below=1 of fh2] (p2) {};
\node[below=1 of p2] (fe2) {};
\node[block, right=2.cm of p2] (Ps2) {\Large $\frac{1}{(\bar{Z}_{e} + Z_{a}) s }$};
\node[splitter, left=1.25 cm of Ps2] (s1) {};
\node[plusminus, right= 1.35cm of Ps2] (sum2) {};
\node[block, right= 1.15cm of sum2] (Cs2) {\small $C(s)$};
\node[plusminus, right= 1. cm of Cs2] (sum22) {};
\node[right= 1.25cm of sum22] (pdes2) {};
\node[below=.5 of p2] (tmp2) {};
\begin{scope}[every path/.style={signal}]
	\node[above right = 1.cm and 1. cm of sum2](fd2) {\textcolor{blue}{$f_{\delta}$}};
	\node[above left = .0cm and .15 cm of fh2](fhl2) {\textcolor{blue}{$f_c$}};
	\node[above left = .0cm and .15 cm of fe2](fel2) {\textcolor{blue}{$f_e$}};
	\node[above left = .0cm and .15 cm of p2](pl2) {\textcolor{blue}{$x_e$}};
	\draw (fh2) -| node[pos=0.9,left] {\scriptsize $-$} (sum2);
	\draw (fe2) -| node[pos=0.9,left] {\scriptsize $-$} (sum2);
	\draw (sum2) -- (Ps2);
	\draw (Cs2) -- node[pos=0.5,below] {\scriptsize $+$} node[pos=0.3,above] {\textcolor{blue}{$f_a$}} (sum2);
	\draw (sum22)--(Cs2);
    \draw (Ps2) -- (s1);
    \draw (s1) -- (p2);
	\draw (s1) -- ++(0cm, -.675cm) -| node[pos=0.8,right ] {\scriptsize $-$} (sum22);
	
	\draw (pdes2) -- node[above right=0cm and 0cm]{\textcolor{blue}{$x_{d}$}} node[pos=0.8,above ] {\scriptsize $+$} (sum22);
	\draw (fd2) -- node[pos=0.7,above ] {\scriptsize $+$} (sum2);
\end{scope}
\draw [color=gray,thick](.50,-.25) rectangle (6.00,-2.75);
\node at (0.75,-.) [above=5mm, right=0mm] {\small $\rm{Position \; \; Controlled \; \; Robot}$};
\end{scope}

\begin{scope}[xshift=6.5cm, yshift=-5.75cm, on grid]
	\node[] (partb) {$\rm{(d)}$};
	\node[below= .4] (fh2) {};
	\node[below=1 of fh2] (p2) {};
	\node[below=1 of p2] (fe2) {};
	\node[block, right=1.5cm of p2] (Ps2) {\Large $\frac{1}{\bar{Z}_{e} s}$};
	\node[plusminus, right=1.5cm of Ps2] (sum2) {};
	\node[block, right=2. of sum2] (Cs2) {\small $P_{s} (s)$};
    \node [splitter, left=1.25cm of Cs2] (s2) {};
	\node[right=1.75 of Cs2] (fdes) {};
	\node[below=.5 of p2] (tmp2) {};
	\begin{scope}[every path/.style={signal}]
		\node[above right = 1.cm and .0 cm of Cs2](fd2) {\textcolor{blue}{$f_{\delta}$}};
		\node[above left = .0cm and .15 cm of fh2](fhl2) {\textcolor{blue}{$f_c$}};
		\node[above left = .0cm and .15 cm of fe2](fel2) {\textcolor{blue}{$f_e$}};
		\node[above left = .0cm and .15 cm of p2](pl2) {\textcolor{blue}{$x_e$}};
		\draw (fh2) -| node[pos=0.9,left ] {\scriptsize $-$} (sum2);
		\draw (fe2) -| node[pos=0.9,left ] {\scriptsize $-$} (sum2);
		\draw (Ps2) -- (p2);
		\draw (sum2) -- (Ps2);
		\draw (Cs2) -- (s2);
        \draw (s2) -- node[pos=0.5,above] {\scriptsize $+$} (sum2);
		\node [below=.75cm of fdes] (springf) {\textcolor{blue}{$f_s$}};
		\draw (s2) |- (springf);
		\draw (fdes) -- node[above right=0cm and 0cm]{\textcolor{blue}{$f_{d}$}} (Cs2);
		\draw (fd2) -- (Cs2);
	\end{scope}
	\draw [color=gray,thick](.5,-.25) rectangle (6.00,-2.75);
	\node at (0.75,-.05) [above=5.mm, right=0mm] {\small $\rm{Force \; \; Controlled \; \; Robot}$};
\end{scope}

\begin{scope}[xshift=7.5cm, yshift=-.2cm]
	\node[plusminus] (fh) {};
    \node[splitter, right=1.5 of fh] (s1) {};
	
	\node[block, below of=fh] (Hs) {$Z_{h} s$};
	
	\node[splitter, below of=Hs] (sp1) {};
	\node[block, right=.75 of sp1, minimum height=3cm, minimum width=1.8cm] (Gs) {\small $\rm{F.C. \; Robot}$};

	\node[block, right= .75 of Gs] (Ks) { $C_{\alpha}(s)$};
	\node[plusminus, right= .75 of Ks] (err) {};
	\node[block, above of=err] (alphas) { $\alpha-1$};
	
	\node[block, below of=sp1] (Es) {$\tilde{Z}_{e} s$};
	
	\node[plusminus, below of=Es] (fe) {};
	\node[left of=fh] (uh) {};
	\node[left of=fe] (ue) {};
	
	\node[left= .25 of uh, above=.5 of uh] (parta) {$\rm{(c)}$};
	\begin{scope}[every path/.style={signal}]
		\draw (Es) -- node[near end,right]{\scriptsize $+$} (fe);
		\draw (fe) -| node[below]{\textcolor{blue}{$f_e$}} (Gs);
        \draw (Gs) -- ++(0.925cm, -1.0cm) -| node[pos= 0.25, below] {\textcolor{blue}{$f_s$}} node[pos= 0.95,right] {\scriptsize $-$} (err);
		\draw (Gs) -- node[pos=1.2, left]{\textcolor{blue}{$x_e$}} (sp1);
		\draw (sp1) -- (Hs);
		\draw (sp1) -- (Es);
		\draw (Hs) -- node[near end,right]{\scriptsize $+$} (fh);
		\draw (fh) -- node[pos= 0.8,above] {\textcolor{blue}{$f_c$}} (s1);
        \draw (s1) -- (Gs);
		\draw (s1) -| (alphas);
		\draw (alphas) -- node[pos= 0.7,right] {\scriptsize $-$} (err);
		\draw (err) -- node[above]{\textcolor{blue}{$- f_{\alpha}$}} (Ks);
		\draw (Ks) -- node[above]{\textcolor{blue}{$f_{d}$}} (Gs);
		\draw (uh) -- node[left=.25 of uh]{\textcolor{blue}{$u_h$}} node[pos= 0.7,above] {\scriptsize $-$}(fh);
		\draw (ue) -- node[left=.25 of ue]{\textcolor{blue}{$f_{ext}$}} node[pos= 0.7,below] {\scriptsize $-$} (fe);
	\end{scope}
	\draw [color=gray,thick](-.6,.5) rectangle (.6,-1.75);
	\node at (-.75,.75) [above=5mm, right=0mm] {\small $\rm{Human}$};
	\draw [color=gray,thick](-.6,-2.25) rectangle (.6,-4.5);
	\node at (-.75,-4.75) [above=5mm, right=0mm] {\small $\rm{Load}$};

\end{scope}
\end{bondgraph}


%% file: fig-fd-fe-cc.pdf_tex
\begingroup%
  \makeatletter%
  \providecommand\color[2][]{%
    \errmessage{(Inkscape) Color is used for the text in Inkscape, but the package 'color.sty' is not loaded}%
    \renewcommand\color[2][]{}%
  }%
  \providecommand\transparent[1]{%
    \errmessage{(Inkscape) Transparency is used (non-zero) for the text in Inkscape, but the package 'transparent.sty' is not loaded}%
    \renewcommand\transparent[1]{}%
  }%
  \providecommand\rotatebox[2]{#2}%
  \ifx\svgwidth\undefined%
    \setlength{\unitlength}{252bp}%
    \ifx\svgscale\undefined%
      \relax%
    \else%
      \setlength{\unitlength}{\unitlength * \real{\svgscale}}%
    \fi%
  \else%
    \setlength{\unitlength}{\svgwidth}%
  \fi%
  \global\let\svgwidth\undefined%
  \global\let\svgscale\undefined%
  \makeatother%
  \begin{picture}(1,0.8)%
    \put(0,0){\includegraphics[width=\unitlength]{fig-fd-fe-cc.pdf}}%
    \put(0.02664871,0.51375794){\rotatebox{90}{\makebox(0,0)[lb]{\smash{$0$}}}}%
    \put(0.02122956,0.56751984){\rotatebox{90}{\makebox(0,0)[lb]{\smash{$20 \cdot \log (\alpha)$}}}}%
    \put(0.48499694,0.70210169){\makebox(0,0)[lb]{\smash{$\omega_{h \mhyphen e}$}}}%
    \put(0.61054325,0.64645089){\makebox(0,0)[lb]{\smash{$\omega_{\alpha h \mhyphen e}$}}}%
    \put(0.74799436,0.64645089){\makebox(0,0)[lb]{\smash{$\omega_{ss \mhyphen a}$}}}%
    \put(0.14270915,0.68275643){\makebox(0,0)[lb]{\smash{Potential Bandwidth}}}%
    \put(0.67411008,0.7097977){\makebox(0,0)[lb]{\smash{Multi-}}}%
    \put(0.64672912,0.67696526){\makebox(0,0)[lb]{\smash{Crossover}}}%
    \put(0.85405889,0.68275643){\makebox(0,0)[lb]{\smash{Noise}}}%
    \put(0.05748948,0.77108333){\makebox(0,0)[lb]{\smash{(a) Magnitude (dB)}}}%
    \put(0.02900486,0.00901424){\rotatebox{90}{\makebox(0,0)[lb]{\smash{$-180$}}}}%
    \put(0.02900486,0.11712217){\rotatebox{90}{\makebox(0,0)[lb]{\smash{$-90$}}}}%
    \put(0.02900486,0.23418968){\rotatebox{90}{\makebox(0,0)[lb]{\smash{$0$}}}}%
    \put(0.48499694,0.21384296){\makebox(0,0)[lb]{\smash{$\omega_{h \mhyphen e}$}}}%
    \put(0.61054325,0.21384296){\makebox(0,0)[lb]{\smash{$\omega_{\alpha h \mhyphen e}$}}}%
    \put(0.74799436,0.21384296){\makebox(0,0)[lb]{\smash{$\omega_{ss \mhyphen a}$}}}%
    \put(0.14270915,0.28075645){\makebox(0,0)[lb]{\smash{Potential Bandwidth}}}%
    \put(0.67411008,0.30779772){\makebox(0,0)[lb]{\smash{Multi-}}}%
    \put(0.64672912,0.27496528){\makebox(0,0)[lb]{\smash{Crossover}}}%
    \put(0.85405889,0.28075645){\makebox(0,0)[lb]{\smash{Noise}}}%
    \put(0.05748948,0.36908333){\makebox(0,0)[lb]{\smash{(b) Phase (deg)}}}%
    \put(0.39668889,0.36813438){\makebox(0,0)[lb]{\smash{\small $P_{\alpha}(s)$}}}%
    \put(0.58954603,0.36813438){\makebox(0,0)[lb]{\smash{\small $C_{\alpha}(s)$}}}%
    \put(0.78240317,0.36813438){\makebox(0,0)[lb]{\smash{\small $P_{\alpha}(s) \cdot C_{\alpha}(s)$}}}%
  \end{picture}%
\endgroup%

%% file: fig-setup.pdf_tex
\begingroup%
  \makeatletter%
  \providecommand\color[2][]{%
    \errmessage{(Inkscape) Color is used for the text in Inkscape, but the package 'color.sty' is not loaded}%
    \renewcommand\color[2][]{}%
  }%
  \providecommand\transparent[1]{%
    \errmessage{(Inkscape) Transparency is used (non-zero) for the text in Inkscape, but the package 'transparent.sty' is not loaded}%
    \renewcommand\transparent[1]{}%
  }%
  \providecommand\rotatebox[2]{#2}%
  \ifx\svgwidth\undefined%
    \setlength{\unitlength}{252bp}%
    \ifx\svgscale\undefined%
      \relax%
    \else%
      \setlength{\unitlength}{\unitlength * \real{\svgscale}}%
    \fi%
  \else%
    \setlength{\unitlength}{\svgwidth}%
  \fi%
  \global\let\svgwidth\undefined%
  \global\let\svgscale\undefined%
  \makeatother%
  \begin{picture}(1,0.32571428)%
    \put(0,0){\includegraphics[width=\unitlength]{fig-setup.pdf}}%
    \put(0.35964753,0.13248971){\color[rgb]{0,0,0}\makebox(0,0)[lb]{\smash{Spring}}}%
    \put(0.3364081,0.2560363){\color[rgb]{0,0,0}\makebox(0,0)[lb]{\smash{Motor}}}%
    \put(0.00299083,0.29028575){\color[rgb]{0,0,0}\makebox(0,0)[lb]{\smash{(a)}}}%
    \put(0.51374788,0.29028575){\color[rgb]{0,0,0}\makebox(0,0)[lb]{\smash{(b)}}}%
    \put(0.51536182,0.15892937){\color[rgb]{0,0,0}\makebox(0,0)[lb]{\smash{Load}}}%
    \put(0.00254094,0.00550094){\color[rgb]{0,0,0}\makebox(0,0)[lb]{\smash{Contact Sensor}}}%
  \end{picture}%
\endgroup%

%% file: fig-xe-fs-ll-00.pdf_tex
\begingroup%
  \makeatletter%
  \providecommand\color[2][]{%
    \errmessage{(Inkscape) Color is used for the text in Inkscape, but the package 'color.sty' is not loaded}%
    \renewcommand\color[2][]{}%
  }%
  \providecommand\transparent[1]{%
    \errmessage{(Inkscape) Transparency is used (non-zero) for the text in Inkscape, but the package 'transparent.sty' is not loaded}%
    \renewcommand\transparent[1]{}%
  }%
  \providecommand\rotatebox[2]{#2}%
  \ifx\svgwidth\undefined%
    \setlength{\unitlength}{252bp}%
    \ifx\svgscale\undefined%
      \relax%
    \else%
      \setlength{\unitlength}{\unitlength * \real{\svgscale}}%
    \fi%
  \else%
    \setlength{\unitlength}{\svgwidth}%
  \fi%
  \global\let\svgwidth\undefined%
  \global\let\svgscale\undefined%
  \makeatother%
  \begin{picture}(1,0.8)%
    \put(0,0){\includegraphics[width=\unitlength]{fig-xe-fs-ll-00.pdf}}%
    \put(0.3884127,0.41354895){\makebox(0,0)[lb]{\smash{$10 ^ {1}$}}}%
    \put(0.83285714,0.41354895){\makebox(0,0)[lb]{\smash{$10 ^ {2}$}}}%
    \put(0.03642857,0.46987188){\makebox(0,0)[lb]{\smash{$0$}}}%
    \put(0.02055556,0.52207107){\makebox(0,0)[lb]{\smash{$20$}}}%
    \put(0.02055556,0.5742703){\makebox(0,0)[lb]{\smash{$40$}}}%
    \put(0.02055556,0.62646947){\makebox(0,0)[lb]{\smash{$60$}}}%
    \put(0.02055556,0.6786687){\makebox(0,0)[lb]{\smash{$80$}}}%
    \put(0.00468254,0.73086792){\makebox(0,0)[lb]{\smash{$100$}}}%
    \put(0.49045778,0.41739567){\makebox(0,0)[lb]{\smash{$\omega \; (rad/s)$}}}%
    \put(0.10111111,0.76869032){\makebox(0,0)[lb]{\smash{(a) Magnitude (dB)}}}%
    \put(0.3884127,0.00754895){\makebox(0,0)[lb]{\smash{$10 ^ {1}$}}}%
    \put(0.83285714,0.00754895){\makebox(0,0)[lb]{\smash{$10 ^ {2}$}}}%
    \put(-0.01323661,0.06387187){\makebox(0,0)[lb]{\smash{$-180$}}}%
    \put(0.00263641,0.12066623){\makebox(0,0)[lb]{\smash{$-90$}}}%
    \put(0.03642857,0.1774606){\makebox(0,0)[lb]{\smash{$0$}}}%
    \put(0.02055556,0.23425497){\makebox(0,0)[lb]{\smash{$90$}}}%
    \put(0.00468254,0.29104934){\makebox(0,0)[lb]{\smash{$180$}}}%
    \put(0.49045778,0.01038459){\makebox(0,0)[lb]{\smash{$\omega \; (rad/s)$}}}%
    \put(0.10111111,0.36269032){\makebox(0,0)[lb]{\smash{(b) Phase (deg)}}}%
    \put(0.50572664,0.36150993){\makebox(0,0)[lb]{\smash{\small Fitted Models}}}%
    \put(0.50662181,0.77026458){\makebox(0,0)[lb]{\smash{\small Exp. IV-A.1}}}%
    \put(0.7713213,0.77026458){\makebox(0,0)[lb]{\smash{\small Exp. IV-A.2}}}%
  \end{picture}%
\endgroup%

%% file: fig-xe-fs-hh-00.pdf_tex
\begingroup%
  \makeatletter%
  \providecommand\color[2][]{%
    \errmessage{(Inkscape) Color is used for the text in Inkscape, but the package 'color.sty' is not loaded}%
    \renewcommand\color[2][]{}%
  }%
  \providecommand\transparent[1]{%
    \errmessage{(Inkscape) Transparency is used (non-zero) for the text in Inkscape, but the package 'transparent.sty' is not loaded}%
    \renewcommand\transparent[1]{}%
  }%
  \providecommand\rotatebox[2]{#2}%
  \ifx\svgwidth\undefined%
    \setlength{\unitlength}{252bp}%
    \ifx\svgscale\undefined%
      \relax%
    \else%
      \setlength{\unitlength}{\unitlength * \real{\svgscale}}%
    \fi%
  \else%
    \setlength{\unitlength}{\svgwidth}%
  \fi%
  \global\let\svgwidth\undefined%
  \global\let\svgscale\undefined%
  \makeatother%
  \begin{picture}(1,0.8)%
    \put(0,0){\includegraphics[width=\unitlength]{fig-xe-fs-hh-00.pdf}}%
    \put(0.07730159,0.4135492){\makebox(0,0)[lb]{\smash{$10 ^ {0}$}}}%
    \put(0.51466107,0.4135492){\makebox(0,0)[lb]{\smash{$10 ^ {1}$}}}%
    \put(0.95202056,0.4135492){\makebox(0,0)[lb]{\smash{$10 ^ {2}$}}}%
    \put(0.03642857,0.46987213){\makebox(0,0)[lb]{\smash{$0$}}}%
    \put(0.02055556,0.52207133){\makebox(0,0)[lb]{\smash{$20$}}}%
    \put(0.02055556,0.57427056){\makebox(0,0)[lb]{\smash{$40$}}}%
    \put(0.02055556,0.62646972){\makebox(0,0)[lb]{\smash{$60$}}}%
    \put(0.02055556,0.67866895){\makebox(0,0)[lb]{\smash{$80$}}}%
    \put(0.00468254,0.73086818){\makebox(0,0)[lb]{\smash{$100$}}}%
    \put(0.58488726,0.41739593){\makebox(0,0)[lb]{\smash{$\omega \; (rad/s)$}}}%
    \put(0.10111111,0.76869057){\makebox(0,0)[lb]{\smash{(a) Magnitude (dB)}}}%
    \put(0.07730159,0.0075492){\makebox(0,0)[lb]{\smash{$10 ^ {0}$}}}%
    \put(0.51466107,0.0075492){\makebox(0,0)[lb]{\smash{$10 ^ {1}$}}}%
    \put(0.95202056,0.0075492){\makebox(0,0)[lb]{\smash{$10 ^ {2}$}}}%
    \put(-0.01323661,0.06387212){\makebox(0,0)[lb]{\smash{$-180$}}}%
    \put(0.00263641,0.12063227){\makebox(0,0)[lb]{\smash{$-90$}}}%
    \put(0.03642857,0.17739243){\makebox(0,0)[lb]{\smash{$0$}}}%
    \put(0.02055556,0.2341526){\makebox(0,0)[lb]{\smash{$90$}}}%
    \put(0.00468254,0.29091275){\makebox(0,0)[lb]{\smash{$180$}}}%
    \put(0.58488726,0.01042475){\makebox(0,0)[lb]{\smash{$\omega \; (rad/s)$}}}%
    \put(0.10111111,0.36269057){\makebox(0,0)[lb]{\smash{(b) Phase (deg)}}}%
    \put(0.50662306,0.77026392){\makebox(0,0)[lb]{\smash{\small Exp. IV-A.3}}}%
    \put(0.77132254,0.77026392){\makebox(0,0)[lb]{\smash{\small Exp. IV-A.4}}}%
    \put(0.50572787,0.36150915){\makebox(0,0)[lb]{\smash{\small Fitted Models}}}%
  \end{picture}%
\endgroup%

%% file: fig-fd-fe-dd-00.pdf_tex
\begingroup%
  \makeatletter%
  \providecommand\color[2][]{%
    \errmessage{(Inkscape) Color is used for the text in Inkscape, but the package 'color.sty' is not loaded}%
    \renewcommand\color[2][]{}%
  }%
  \providecommand\transparent[1]{%
    \errmessage{(Inkscape) Transparency is used (non-zero) for the text in Inkscape, but the package 'transparent.sty' is not loaded}%
    \renewcommand\transparent[1]{}%
  }%
  \providecommand\rotatebox[2]{#2}%
  \ifx\svgwidth\undefined%
    \setlength{\unitlength}{252bp}%
    \ifx\svgscale\undefined%
      \relax%
    \else%
      \setlength{\unitlength}{\unitlength * \real{\svgscale}}%
    \fi%
  \else%
    \setlength{\unitlength}{\svgwidth}%
  \fi%
  \global\let\svgwidth\undefined%
  \global\let\svgscale\undefined%
  \makeatother%
  \begin{picture}(1,0.8)%
    \put(0,0){\includegraphics[width=\unitlength]{fig-fd-fe-dd-00.pdf}}%
    \put(0.07730159,0.41274895){\makebox(0,0)[lb]{\smash{$10 ^ {0}$}}}%
    \put(0.463775,0.41274895){\makebox(0,0)[lb]{\smash{$10 ^ {1}$}}}%
    \put(0.85024845,0.41274895){\makebox(0,0)[lb]{\smash{$10 ^ {2}$}}}%
    \put(0.00263641,0.47914096){\makebox(0,0)[lb]{\smash{$-20$}}}%
    \put(0.03642857,0.57983171){\makebox(0,0)[lb]{\smash{$0$}}}%
    \put(0.02055556,0.68052252){\makebox(0,0)[lb]{\smash{$20$}}}%
    \put(0.54338345,0.41639567){\makebox(0,0)[lb]{\smash{$\omega \; (rad/s)$}}}%
    \put(0.10111111,0.76869032){\makebox(0,0)[lb]{\smash{(a) Magnitude (dB)}}}%
    \put(0.07730159,0.00594895){\makebox(0,0)[lb]{\smash{$10 ^ {0}$}}}%
    \put(0.463775,0.00594895){\makebox(0,0)[lb]{\smash{$10 ^ {1}$}}}%
    \put(0.85024845,0.00594895){\makebox(0,0)[lb]{\smash{$10 ^ {2}$}}}%
    \put(-0.01323661,0.07856231){\makebox(0,0)[lb]{\smash{$-180$}}}%
    \put(0.00263641,0.19572421){\makebox(0,0)[lb]{\smash{$-90$}}}%
    \put(0.03642857,0.31288609){\makebox(0,0)[lb]{\smash{$0$}}}%
    \put(0.54338345,0.014529){\makebox(0,0)[lb]{\smash{$\omega \; (rad/s)$}}}%
    \put(0.10111111,0.36189032){\makebox(0,0)[lb]{\smash{(b) Phase (deg)}}}%
    \put(0.43319683,0.36410929){\makebox(0,0)[lb]{\smash{\small $P_{\alpha}(s)$}}}%
    \put(0.61018095,0.36410929){\makebox(0,0)[lb]{\smash{\small $C_{\alpha}(s)$}}}%
    \put(0.78716508,0.36410929){\makebox(0,0)[lb]{\smash{\small $P_{\alpha}(s) \cdot C_{\alpha}(s)$}}}%
  \end{picture}%
\endgroup%

%% file: fig-fs-fc-p-setup.pdf_tex
\begingroup%
  \makeatletter%
  \providecommand\color[2][]{%
    \errmessage{(Inkscape) Color is used for the text in Inkscape, but the package 'color.sty' is not loaded}%
    \renewcommand\color[2][]{}%
  }%
  \providecommand\transparent[1]{%
    \errmessage{(Inkscape) Transparency is used (non-zero) for the text in Inkscape, but the package 'transparent.sty' is not loaded}%
    \renewcommand\transparent[1]{}%
  }%
  \providecommand\rotatebox[2]{#2}%
  \ifx\svgwidth\undefined%
    \setlength{\unitlength}{252bp}%
    \ifx\svgscale\undefined%
      \relax%
    \else%
      \setlength{\unitlength}{\unitlength * \real{\svgscale}}%
    \fi%
  \else%
    \setlength{\unitlength}{\svgwidth}%
  \fi%
  \global\let\svgwidth\undefined%
  \global\let\svgscale\undefined%
  \makeatother%
  \begin{picture}(1,0.32571428)%
    \put(0,0){\includegraphics[width=\unitlength]{fig-fs-fc-p-setup.pdf}}%
    \put(0.00210317,0.29420536){\color[rgb]{0,0,0}\makebox(0,0)[lb]{\smash{(a)}}}%
    \put(0.51210317,0.29420536){\color[rgb]{0,0,0}\makebox(0,0)[lb]{\smash{(b)}}}%
    \put(0.20714533,0.26534329){\color[rgb]{0,0,0}\makebox(0,0)[lb]{\smash{Open Hand}}}%
    \put(0.71285962,0.26534329){\color[rgb]{0,0,0}\makebox(0,0)[lb]{\smash{Closed Hand}}}%
    \put(0.02643841,0.14328762){\color[rgb]{0,0,0}\rotatebox{-51.94187816}{\makebox(0,0)[lb]{\smash{\small Trigger}}}}%
    \put(0.54093133,0.14518632){\color[rgb]{0,0,0}\rotatebox{-51.94187816}{\makebox(0,0)[lb]{\smash{\small Trigger}}}}%
  \end{picture}%
\endgroup%

%% file: fig-fs-fc-p.pdf_tex
\begingroup%
  \makeatletter%
  \providecommand\color[2][]{%
    \errmessage{(Inkscape) Color is used for the text in Inkscape, but the package 'color.sty' is not loaded}%
    \renewcommand\color[2][]{}%
  }%
  \providecommand\transparent[1]{%
    \errmessage{(Inkscape) Transparency is used (non-zero) for the text in Inkscape, but the package 'transparent.sty' is not loaded}%
    \renewcommand\transparent[1]{}%
  }%
  \providecommand\rotatebox[2]{#2}%
  \ifx\svgwidth\undefined%
    \setlength{\unitlength}{252bp}%
    \ifx\svgscale\undefined%
      \relax%
    \else%
      \setlength{\unitlength}{\unitlength * \real{\svgscale}}%
    \fi%
  \else%
    \setlength{\unitlength}{\svgwidth}%
  \fi%
  \global\let\svgwidth\undefined%
  \global\let\svgscale\undefined%
  \makeatother%
  \begin{picture}(1,0.6)%
    \put(0,0){\includegraphics[width=\unitlength]{fig-fs-fc-p.pdf}}%
    \put(0.08746031,0.30486593){\makebox(0,0)[lb]{\smash{$0$}}}%
    \put(0.22232803,0.30486593){\makebox(0,0)[lb]{\smash{$1$}}}%
    \put(0.35719576,0.30486593){\makebox(0,0)[lb]{\smash{$2$}}}%
    \put(0.49206348,0.30486593){\makebox(0,0)[lb]{\smash{$3$}}}%
    \put(0.62693122,0.30486593){\makebox(0,0)[lb]{\smash{$4$}}}%
    \put(0.76179892,0.30486593){\makebox(0,0)[lb]{\smash{$5$}}}%
    \put(-0.00307789,0.35201375){\makebox(0,0)[lb]{\smash{$-10$}}}%
    \put(0.01279513,0.4119319){\makebox(0,0)[lb]{\smash{$-5$}}}%
    \put(0.03071428,0.47185007){\makebox(0,0)[lb]{\smash{$0$}}}%
    \put(-0.00617197,0.51868089){\makebox(0,0)[lb]{\smash{\footnotesize $\rm{\tau (Nm)}$}}}%
    \put(0.85857936,0.31086023){\makebox(0,0)[lb]{\smash{$\rm{t \; (s)}$}}}%
    \put(0.91886773,0.51868089){\makebox(0,0)[lb]{\smash{\footnotesize $\rm{\theta_e (rad)}$}}}%
    \put(0.09539682,0.56959063){\makebox(0,0)[lb]{\smash{$\rm{(a) \; 10 \; lb \; Load \; \mhyphen \; Open \; Hand}$}}}%
    \put(0.64198412,0.57158252){\makebox(0,0)[lb]{\smash{$\theta_{e}$}}}%
    \put(0.78642856,0.5715825){\makebox(0,0)[lb]{\smash{$\tau_c$}}}%
    \put(0.93087301,0.5715825){\makebox(0,0)[lb]{\smash{$\hat{\tau}_c$}}}%
    \put(0.95341273,0.35201375){\makebox(0,0)[lb]{\smash{$1.8$}}}%
    \put(0.95341273,0.4119319){\makebox(0,0)[lb]{\smash{$2.0$}}}%
    \put(0.95341273,0.47185007){\makebox(0,0)[lb]{\smash{$2.2$}}}%
    \put(0.08746031,0.00291405){\makebox(0,0)[lb]{\smash{$0$}}}%
    \put(0.22232803,0.00291405){\makebox(0,0)[lb]{\smash{$1$}}}%
    \put(0.35719576,0.00291405){\makebox(0,0)[lb]{\smash{$2$}}}%
    \put(0.49206348,0.00291405){\makebox(0,0)[lb]{\smash{$3$}}}%
    \put(0.62693122,0.00291405){\makebox(0,0)[lb]{\smash{$4$}}}%
    \put(0.76179892,0.00291405){\makebox(0,0)[lb]{\smash{$5$}}}%
    \put(-0.00307789,0.05006187){\makebox(0,0)[lb]{\smash{$-10$}}}%
    \put(0.01279513,0.10998002){\makebox(0,0)[lb]{\smash{$-5$}}}%
    \put(0.03071428,0.16989818){\makebox(0,0)[lb]{\smash{$0$}}}%
    \put(-0.00617197,0.21672903){\makebox(0,0)[lb]{\smash{\footnotesize $\rm{\tau (Nm)}$}}}%
    \put(0.85857936,0.00890834){\makebox(0,0)[lb]{\smash{$\rm{t \; (s)}$}}}%
    \put(0.91886773,0.21672903){\makebox(0,0)[lb]{\smash{\footnotesize $\rm{\theta_e (rad)}$}}}%
    \put(0.09539682,0.26763875){\makebox(0,0)[lb]{\smash{$\rm{(b) \; 10 \; lb \; Load \; \mhyphen \; Closed\; Hand}$}}}%
    \put(0.95341273,0.05006187){\makebox(0,0)[lb]{\smash{$1.8$}}}%
    \put(0.95341273,0.10998002){\makebox(0,0)[lb]{\smash{$2.0$}}}%
    \put(0.95341273,0.16989818){\makebox(0,0)[lb]{\smash{$2.2$}}}%
  \end{picture}%
\endgroup%

%% file: fig-fd-fe-aa-00.pdf_tex
\begingroup%
  \makeatletter%
  \providecommand\color[2][]{%
    \errmessage{(Inkscape) Color is used for the text in Inkscape, but the package 'color.sty' is not loaded}%
    \renewcommand\color[2][]{}%
  }%
  \providecommand\transparent[1]{%
    \errmessage{(Inkscape) Transparency is used (non-zero) for the text in Inkscape, but the package 'transparent.sty' is not loaded}%
    \renewcommand\transparent[1]{}%
  }%
  \providecommand\rotatebox[2]{#2}%
  \ifx\svgwidth\undefined%
    \setlength{\unitlength}{252bp}%
    \ifx\svgscale\undefined%
      \relax%
    \else%
      \setlength{\unitlength}{\unitlength * \real{\svgscale}}%
    \fi%
  \else%
    \setlength{\unitlength}{\svgwidth}%
  \fi%
  \global\let\svgwidth\undefined%
  \global\let\svgscale\undefined%
  \makeatother%
  \begin{picture}(1,0.8)%
    \put(0,0){\includegraphics[width=\unitlength]{fig-fd-fe-aa-00.pdf}}%
    \put(0.07730159,0.41274895){\makebox(0,0)[lb]{\smash{$10 ^ {0}$}}}%
    \put(0.463775,0.41274895){\makebox(0,0)[lb]{\smash{$10 ^ {1}$}}}%
    \put(0.85024845,0.41274895){\makebox(0,0)[lb]{\smash{$10 ^ {2}$}}}%
    \put(0.00263641,0.47914096){\makebox(0,0)[lb]{\smash{$-20$}}}%
    \put(0.03642857,0.57983171){\makebox(0,0)[lb]{\smash{$0$}}}%
    \put(0.02055556,0.68052252){\makebox(0,0)[lb]{\smash{$20$}}}%
    \put(0.54338345,0.41639567){\makebox(0,0)[lb]{\smash{$\omega \; (rad/s)$}}}%
    \put(0.10111111,0.76869032){\makebox(0,0)[lb]{\smash{(a) Magnitude (dB)}}}%
    \put(0.07730159,0.00594895){\makebox(0,0)[lb]{\smash{$10 ^ {0}$}}}%
    \put(0.463775,0.00594895){\makebox(0,0)[lb]{\smash{$10 ^ {1}$}}}%
    \put(0.85024845,0.00594895){\makebox(0,0)[lb]{\smash{$10 ^ {2}$}}}%
    \put(-0.01323661,0.07417169){\makebox(0,0)[lb]{\smash{$-180$}}}%
    \put(0.00263641,0.19350626){\makebox(0,0)[lb]{\smash{$-90$}}}%
    \put(0.03642857,0.31284086){\makebox(0,0)[lb]{\smash{$0$}}}%
    \put(0.54338345,0.00871405){\makebox(0,0)[lb]{\smash{$\omega \; (rad/s)$}}}%
    \put(0.10111111,0.36189032){\makebox(0,0)[lb]{\smash{(b) Phase (deg)}}}%
    \put(0.43319683,0.36410929){\makebox(0,0)[lb]{\smash{\small $P_{\alpha}(s)$}}}%
    \put(0.61018095,0.36410929){\makebox(0,0)[lb]{\smash{\small $C_{\alpha}(s)$}}}%
    \put(0.78716508,0.36410929){\makebox(0,0)[lb]{\smash{\small $P_{\alpha}(s) \cdot C_{\alpha}(s)$}}}%
  \end{picture}%
\endgroup%

%% file: fig-fs-fc-s-f.pdf_tex
\begingroup%
  \makeatletter%
  \providecommand\color[2][]{%
    \errmessage{(Inkscape) Color is used for the text in Inkscape, but the package 'color.sty' is not loaded}%
    \renewcommand\color[2][]{}%
  }%
  \providecommand\transparent[1]{%
    \errmessage{(Inkscape) Transparency is used (non-zero) for the text in Inkscape, but the package 'transparent.sty' is not loaded}%
    \renewcommand\transparent[1]{}%
  }%
  \providecommand\rotatebox[2]{#2}%
  \ifx\svgwidth\undefined%
    \setlength{\unitlength}{518.4bp}%
    \ifx\svgscale\undefined%
      \relax%
    \else%
      \setlength{\unitlength}{\unitlength * \real{\svgscale}}%
    \fi%
  \else%
    \setlength{\unitlength}{\svgwidth}%
  \fi%
  \global\let\svgwidth\undefined%
  \global\let\svgscale\undefined%
  \makeatother%
  \begin{picture}(1,0.48611111)%
    \put(0,0){\includegraphics[width=\unitlength]{fig-fs-fc-s-f.pdf}}%
    \put(0.03980179,0.12450716){\makebox(0,0)[lb]{\smash{$0$}}}%
    \put(0.10695392,0.12450716){\makebox(0,0)[lb]{\smash{$2$}}}%
    \put(0.17410606,0.12450716){\makebox(0,0)[lb]{\smash{$4$}}}%
    \put(0.24125819,0.12450716){\makebox(0,0)[lb]{\smash{$6$}}}%
    \put(0.30841033,0.12450716){\makebox(0,0)[lb]{\smash{$8$}}}%
    \put(0.37170443,0.12450716){\makebox(0,0)[lb]{\smash{$10$}}}%
    \put(-0.00420984,0.14742623){\makebox(0,0)[lb]{\smash{$-10$}}}%
    \put(0.01221691,0.16278573){\makebox(0,0)[lb]{\smash{$0$}}}%
    \put(0.00450086,0.17814523){\makebox(0,0)[lb]{\smash{$10$}}}%
    \put(0.00450086,0.19350473){\makebox(0,0)[lb]{\smash{$20$}}}%
    \put(-0.00619262,0.21142866){\makebox(0,0)[lb]{\smash{\footnotesize $\rm{\tau (Nm)}$}}}%
    \put(0.42396122,0.1244522){\makebox(0,0)[lb]{\smash{$\rm{t \; (s)}$}}}%
    \put(0.45387216,0.21142866){\makebox(0,0)[lb]{\smash{\footnotesize $\rm{\theta_e (rad)}$}}}%
    \put(0.04365981,0.22725012){\makebox(0,0)[lb]{\smash{$\rm{(c) \; 10 \; lb \; Load \; \mhyphen\; Open \; Hand}$}}}%
    \put(0.47029948,0.14742623){\makebox(0,0)[lb]{\smash{$1.5$}}}%
    \put(0.47029948,0.16790556){\makebox(0,0)[lb]{\smash{$2.0$}}}%
    \put(0.47029948,0.1883849){\makebox(0,0)[lb]{\smash{$2.5$}}}%
    \put(0.03980179,0.00338552){\makebox(0,0)[lb]{\smash{$0$}}}%
    \put(0.10695392,0.00338552){\makebox(0,0)[lb]{\smash{$2$}}}%
    \put(0.17410606,0.00338552){\makebox(0,0)[lb]{\smash{$4$}}}%
    \put(0.24125819,0.00338552){\makebox(0,0)[lb]{\smash{$6$}}}%
    \put(0.30841033,0.00338552){\makebox(0,0)[lb]{\smash{$8$}}}%
    \put(0.37170443,0.00338552){\makebox(0,0)[lb]{\smash{$10$}}}%
    \put(-0.00420984,0.0263046){\makebox(0,0)[lb]{\smash{$-10$}}}%
    \put(0.01221691,0.0416641){\makebox(0,0)[lb]{\smash{$0$}}}%
    \put(0.00450086,0.0570236){\makebox(0,0)[lb]{\smash{$10$}}}%
    \put(0.00450086,0.07238309){\makebox(0,0)[lb]{\smash{$20$}}}%
    \put(-0.00619262,0.09030703){\makebox(0,0)[lb]{\smash{\footnotesize $\rm{\tau (Nm)}$}}}%
    \put(0.42396122,0.00333056){\makebox(0,0)[lb]{\smash{$\rm{t \; (s)}$}}}%
    \put(0.45387216,0.09030703){\makebox(0,0)[lb]{\smash{\footnotesize $\rm{\theta_e (rad)}$}}}%
    \put(0.04365981,0.10612849){\makebox(0,0)[lb]{\smash{$\rm{(d) \; 10 \; lb \; Load \; \mhyphen\; Closed\; Hand}$}}}%
    \put(0.47029948,0.0263046){\makebox(0,0)[lb]{\smash{$1.5$}}}%
    \put(0.47029948,0.04678393){\makebox(0,0)[lb]{\smash{$2.0$}}}%
    \put(0.47029948,0.06726326){\makebox(0,0)[lb]{\smash{$2.5$}}}%
    \put(0.03980179,0.36675043){\makebox(0,0)[lb]{\smash{$0$}}}%
    \put(0.10695392,0.36675043){\makebox(0,0)[lb]{\smash{$2$}}}%
    \put(0.17410606,0.36675043){\makebox(0,0)[lb]{\smash{$4$}}}%
    \put(0.24125819,0.36675043){\makebox(0,0)[lb]{\smash{$6$}}}%
    \put(0.30841033,0.36675043){\makebox(0,0)[lb]{\smash{$8$}}}%
    \put(0.37170443,0.36675043){\makebox(0,0)[lb]{\smash{$10$}}}%
    \put(0.00350621,0.3896695){\makebox(0,0)[lb]{\smash{$-2$}}}%
    \put(0.01221691,0.41014886){\makebox(0,0)[lb]{\smash{$0$}}}%
    \put(0.01221691,0.43062817){\makebox(0,0)[lb]{\smash{$2$}}}%
    \put(-0.00619262,0.451624){\makebox(0,0)[lb]{\smash{\footnotesize $\rm{\tau (Nm)}$}}}%
    \put(0.42396122,0.36823141){\makebox(0,0)[lb]{\smash{$\rm{t \; (s)}$}}}%
    \put(0.45387216,0.451624){\makebox(0,0)[lb]{\smash{\footnotesize $\rm{\theta_e (rad)}$}}}%
    \put(0.04365981,0.46949341){\makebox(0,0)[lb]{\smash{$\rm{(a) \; 0 \; Load \; \mhyphen\; Open \; Hand}$}}}%
    \put(0.28846884,0.47046169){\makebox(0,0)[lb]{\smash{$\theta_{e}$}}}%
    \put(0.36659384,0.47046168){\makebox(0,0)[lb]{\smash{$-\tau_c$}}}%
    \put(0.46015094,0.47046168){\makebox(0,0)[lb]{\smash{$\tau_s$}}}%
    \put(0.47029948,0.3896695){\makebox(0,0)[lb]{\smash{$1.5$}}}%
    \put(0.47029948,0.41014886){\makebox(0,0)[lb]{\smash{$2.0$}}}%
    \put(0.47029948,0.43062817){\makebox(0,0)[lb]{\smash{$2.5$}}}%
    \put(0.03980179,0.24562879){\makebox(0,0)[lb]{\smash{$0$}}}%
    \put(0.10695392,0.24562879){\makebox(0,0)[lb]{\smash{$2$}}}%
    \put(0.17410606,0.24562879){\makebox(0,0)[lb]{\smash{$4$}}}%
    \put(0.24125819,0.24562879){\makebox(0,0)[lb]{\smash{$6$}}}%
    \put(0.30841033,0.24562879){\makebox(0,0)[lb]{\smash{$8$}}}%
    \put(0.37170443,0.24562879){\makebox(0,0)[lb]{\smash{$10$}}}%
    \put(0.00350621,0.26854786){\makebox(0,0)[lb]{\smash{$-2$}}}%
    \put(0.01221691,0.28902722){\makebox(0,0)[lb]{\smash{$0$}}}%
    \put(0.01221691,0.30950652){\makebox(0,0)[lb]{\smash{$2$}}}%
    \put(-0.00619262,0.33050237){\makebox(0,0)[lb]{\smash{\footnotesize $\rm{\tau (Nm)}$}}}%
    \put(0.42396122,0.24710979){\makebox(0,0)[lb]{\smash{$\rm{t \; (s)}$}}}%
    \put(0.45387216,0.33050237){\makebox(0,0)[lb]{\smash{\footnotesize $\rm{\theta_e (rad)}$}}}%
    \put(0.04365981,0.34837177){\makebox(0,0)[lb]{\smash{$\rm{(b) \; 0\; Load \; \mhyphen\; Closed\; Hand}$}}}%
    \put(0.47029948,0.26854786){\makebox(0,0)[lb]{\smash{$1.5$}}}%
    \put(0.47029948,0.28902722){\makebox(0,0)[lb]{\smash{$2.0$}}}%
    \put(0.47029948,0.30950652){\makebox(0,0)[lb]{\smash{$2.5$}}}%
    \put(0.56549623,0.12450716){\makebox(0,0)[lb]{\smash{$0$}}}%
    \put(0.63264837,0.12450716){\makebox(0,0)[lb]{\smash{$2$}}}%
    \put(0.6998005,0.12450716){\makebox(0,0)[lb]{\smash{$4$}}}%
    \put(0.76695264,0.12450716){\makebox(0,0)[lb]{\smash{$6$}}}%
    \put(0.83410478,0.12450716){\makebox(0,0)[lb]{\smash{$8$}}}%
    \put(0.89739888,0.12450716){\makebox(0,0)[lb]{\smash{$10$}}}%
    \put(0.51839819,0.14742623){\makebox(0,0)[lb]{\smash{$-10$}}}%
    \put(0.53482494,0.16278573){\makebox(0,0)[lb]{\smash{$0$}}}%
    \put(0.52710889,0.17814523){\makebox(0,0)[lb]{\smash{$10$}}}%
    \put(0.52710889,0.19350473){\makebox(0,0)[lb]{\smash{$20$}}}%
    \put(0.51641541,0.21142866){\makebox(0,0)[lb]{\smash{\footnotesize $\rm{\tau (Nm)}$}}}%
    \put(0.94965567,0.1244522){\makebox(0,0)[lb]{\smash{$\rm{t \; (s)}$}}}%
    \put(0.9795666,0.21142866){\makebox(0,0)[lb]{\smash{\footnotesize $\rm{\theta_e (rad)}$}}}%
    \put(0.56935426,0.22725012){\makebox(0,0)[lb]{\smash{$\rm{(g) \; 10 \; lb \; Load \; \mhyphen \; Open \; Hand}$}}}%
    \put(0.99599392,0.14742623){\makebox(0,0)[lb]{\smash{$1.5$}}}%
    \put(0.99599392,0.16790556){\makebox(0,0)[lb]{\smash{$2.0$}}}%
    \put(0.99599392,0.1883849){\makebox(0,0)[lb]{\smash{$2.5$}}}%
    \put(0.56549623,0.00338552){\makebox(0,0)[lb]{\smash{$0$}}}%
    \put(0.63264837,0.00338552){\makebox(0,0)[lb]{\smash{$2$}}}%
    \put(0.6998005,0.00338552){\makebox(0,0)[lb]{\smash{$4$}}}%
    \put(0.76695264,0.00338552){\makebox(0,0)[lb]{\smash{$6$}}}%
    \put(0.83410478,0.00338552){\makebox(0,0)[lb]{\smash{$8$}}}%
    \put(0.89739888,0.00338552){\makebox(0,0)[lb]{\smash{$10$}}}%
    \put(0.52148461,0.0263046){\makebox(0,0)[lb]{\smash{$-10$}}}%
    \put(0.53791136,0.0416641){\makebox(0,0)[lb]{\smash{$0$}}}%
    \put(0.53019531,0.0570236){\makebox(0,0)[lb]{\smash{$10$}}}%
    \put(0.53019531,0.07238309){\makebox(0,0)[lb]{\smash{$20$}}}%
    \put(0.51641541,0.09030703){\makebox(0,0)[lb]{\smash{\footnotesize $\rm{\tau (Nm)}$}}}%
    \put(0.94965567,0.00333056){\makebox(0,0)[lb]{\smash{$\rm{t \; (s)}$}}}%
    \put(0.9795666,0.09030703){\makebox(0,0)[lb]{\smash{\footnotesize $\rm{\theta_e (rad)}$}}}%
    \put(0.56935426,0.10612849){\makebox(0,0)[lb]{\smash{$\rm{(h) \; 10 \; lb \; Load \; \mhyphen\; Closed\; Hand}$}}}%
    \put(0.99599392,0.0263046){\makebox(0,0)[lb]{\smash{$1.5$}}}%
    \put(0.99599392,0.04678393){\makebox(0,0)[lb]{\smash{$2.0$}}}%
    \put(0.99599392,0.06726326){\makebox(0,0)[lb]{\smash{$2.5$}}}%
    \put(0.56549623,0.36675043){\makebox(0,0)[lb]{\smash{$0$}}}%
    \put(0.63264837,0.36675043){\makebox(0,0)[lb]{\smash{$2$}}}%
    \put(0.6998005,0.36675043){\makebox(0,0)[lb]{\smash{$4$}}}%
    \put(0.76695264,0.36675043){\makebox(0,0)[lb]{\smash{$6$}}}%
    \put(0.83410478,0.36675043){\makebox(0,0)[lb]{\smash{$8$}}}%
    \put(0.89739888,0.36675043){\makebox(0,0)[lb]{\smash{$10$}}}%
    \put(0.52611424,0.3896695){\makebox(0,0)[lb]{\smash{$-2$}}}%
    \put(0.53482494,0.41014886){\makebox(0,0)[lb]{\smash{$0$}}}%
    \put(0.53482494,0.43062817){\makebox(0,0)[lb]{\smash{$2$}}}%
    \put(0.51641541,0.451624){\makebox(0,0)[lb]{\smash{\footnotesize $\rm{\tau (Nm)}$}}}%
    \put(0.94965567,0.36823141){\makebox(0,0)[lb]{\smash{$\rm{t \; (s)}$}}}%
    \put(0.9795666,0.451624){\makebox(0,0)[lb]{\smash{\footnotesize $\rm{\theta_e (rad)}$}}}%
    \put(0.56935426,0.46949341){\makebox(0,0)[lb]{\smash{$\rm{(e) \; 0 \; Load \; \mhyphen\; Open \; Hand}$}}}%
    \put(0.99599392,0.3896695){\makebox(0,0)[lb]{\smash{$1.5$}}}%
    \put(0.99599392,0.41014886){\makebox(0,0)[lb]{\smash{$2.0$}}}%
    \put(0.99599392,0.43062817){\makebox(0,0)[lb]{\smash{$2.5$}}}%
    \put(0.56549623,0.24562879){\makebox(0,0)[lb]{\smash{$0$}}}%
    \put(0.63264837,0.24562879){\makebox(0,0)[lb]{\smash{$2$}}}%
    \put(0.6998005,0.24562879){\makebox(0,0)[lb]{\smash{$4$}}}%
    \put(0.76695264,0.24562879){\makebox(0,0)[lb]{\smash{$6$}}}%
    \put(0.83410478,0.24562879){\makebox(0,0)[lb]{\smash{$8$}}}%
    \put(0.89739888,0.24562879){\makebox(0,0)[lb]{\smash{$10$}}}%
    \put(0.52611424,0.26854786){\makebox(0,0)[lb]{\smash{$-2$}}}%
    \put(0.53482494,0.28902722){\makebox(0,0)[lb]{\smash{$0$}}}%
    \put(0.53482494,0.30950652){\makebox(0,0)[lb]{\smash{$2$}}}%
    \put(0.51641541,0.33050237){\makebox(0,0)[lb]{\smash{\footnotesize $\rm{\tau (Nm)}$}}}%
    \put(0.94965567,0.24710979){\makebox(0,0)[lb]{\smash{$\rm{t \; (s)}$}}}%
    \put(0.9795666,0.33050237){\makebox(0,0)[lb]{\smash{\footnotesize $\rm{\theta_e (rad)}$}}}%
    \put(0.56935426,0.34837177){\makebox(0,0)[lb]{\smash{$\rm{(f) \; 0\; Load \; \mhyphen\; Closed\; Hand}$}}}%
    \put(0.99599392,0.26854786){\makebox(0,0)[lb]{\smash{$1.5$}}}%
    \put(0.99599392,0.28902722){\makebox(0,0)[lb]{\smash{$2.0$}}}%
    \put(0.99599392,0.30950652){\makebox(0,0)[lb]{\smash{$2.5$}}}%
  \end{picture}%
\endgroup%